\documentclass[lettersize,journal]{IEEEtran}
\usepackage{graphicx}

\usepackage{pgffor}

\ifCLASSOPTIONcompsoc
\usepackage[nocompress]{cite}
\else
\usepackage{cite}
\fi
\usepackage{hyperref}
\usepackage{pgf-pie}
\usepackage{graphicx}
\usepackage{blkarray}
\usepackage{standalone}
\usepackage{amsmath}
\usepackage{xcolor} 
\usepackage{subfig}
\usepackage{pgfplots}
\usepackage{pgfplotstable}
\usepackage{adjustbox}
\usepackage{booktabs}
\usepackage{csquotes}
\usepackage{multirow}
\usepackage[figuresright]{rotating}
\usepackage[most]{tcolorbox}
\usepackage{colortbl}
\usepackage{xcolor}
\usepackage{balance}
\usepackage{url}
\usepackage{ifxetex,ifluatex}
\usepackage{etoolbox}
\usepackage{tikz}
\usetikzlibrary{patterns}
\usepackage{enumitem}
\usepackage{amsthm}

\usepackage{framed}
\usepackage{bm}
\definecolor{RedLine}{RGB}{200,0,0}
\DeclareMathAlphabet\mathbfcal{OMS}{cmsy}{b}{n}
\pgfplotsset{compat=newest}

\usepackage{amssymb}
\usepackage{pifont}
\usepackage{amsmath}

\definecolor{steel}{rgb}{0.1, 0.2, 0.8} 
\definecolor{lightblue}{rgb}{0.4, 0.6, 1.0}
\definecolor{mediumblue}{rgb}{0.2, 0.4, 0.8}
\definecolor{lightgray}{rgb}{0.8, 0.8, 0.8}
\definecolor{lightred}{rgb}{1.0, 0.6, 0.6}
\definecolor{mediumred}{rgb}{0.9, 0.3, 0.3}
\definecolor{darkred}{rgb}{0.7, 0.1, 0.1}
   
\newtcolorbox{quotebox}{colback=black!10,boxrule=0.4pt,colframe=black,fonttitle=\bfseries,top=2pt,bottom=2pt}

\newcommand{\keybox}[1]{
\begin{tcolorbox}[colback=white,leftrule=1mm,toprule=0mm,bottomrule=0mm,left=1pt,right=2pt,top=2pt,bottom=2pt]
\em #1
\end{tcolorbox}
}

\newcounter{findingcount}
\newcounter{insightcount}
\usepackage{algorithm2e}
\usepackage{xcolor}

\SetAlFnt{\footnotesize}

\SetAlCapSty{xAlCapSty}

\SetCommentSty{xCommentSty}

\SetNlSty{mynlfont}{}{} 

\LinesNumbered

\SetSideCommentRight

\DontPrintSemicolon

\RestyleAlgo{algoruled}

\usepackage[autolanguage]{numprint}

\newcommand*\np[2][z]{
\ifx z#1%
$\numprint{#2}$%
\else%
$\numprint[#1]{#2}$%
\fi\xspace%
}

\usepackage{fp}
\newcommand{\ShowAbsoluteNumber}[1]{%
\ifnum #1<10%
{\hspace*{0pt}#1}%
\else%
\ifnum #1<100%
{\hspace*{0pt}#1}%
\else%
\ifnum #1<1000%
{\hspace*{0pt}#1}%
\else%
{\numprint{#1}}%
\fi%
\fi%
\fi%
}

\newcommand{\ShowPercentage}[2]{%
\FPeval\percentage{round(#1/#2*100,0)}%
\FPeval\percentageOneDecimal{round(#1/#2*100,1)}%
\ifnum \percentage=0%
{\np[\%]{\FPprint{percentageOneDecimal}}}%
\else%
\ifnum \percentage<10%
{\np[\%]{\FPprint{percentageOneDecimal}}}%
\else%
{\np[\%]{\FPprint{percentageOneDecimal}}}%
\fi%
\fi%
\xspace
}

\newcommand{\ShowPercentageTwo}[2]{%
\FPeval\percentagetwo{round(#1/#2*100,0)}%
\FPeval\percentageTwoDecimal{round(#1/#2*100,2)}%
\ifnum \percentagetwo=0%
{\np[\%]{\FPprint{percentageTwoDecimal}}}%
\else%
\ifnum \percentagetwo<10%
{\np[\%]{\FPprint{percentageTwoDecimal}}}%
\else%
{\np[\%]{\FPprint{percentageTwoDecimal}}}%
\fi%
\fi%
\xspace
}

\newlength\BARSIZE  
\newcommand{\inlinechart}[2]{%
\FPeval{\BLACKBARSIZE}{#1/#2}\textcolor{black!80}{\rule{\BLACKBARSIZE\BARSIZE}{1.6ex}}%
\FPeval{\BLACKBARSIZE}{1 - (#1/#2)}\textcolor{black!10}{\rule{\BLACKBARSIZE\BARSIZE}{1.6ex}}%
}

\newcommand{\pinlinechart}[3]{%
\FPeval{\BLACKBARSIZE}{#1/#2}\textcolor{gree!#3}{\rule{\BLACKBARSIZE\BARSIZE}{1.6ex}}%
\FPeval{\BLACKBARSIZE}{1 - (#1/#2)}\textcolor{black!10}{\rule{\BLACKBARSIZE\BARSIZE}{1.6ex}}%
}

\newcommand{\ninlinechart}[3]{%
\FPeval{\BLACKBARSIZE}{1 - (#1/#2)}\textcolor{black!10}{\rule{\BLACKBARSIZE\BARSIZE}{1.6ex}}%
\FPeval{\BLACKBARSIZE}{#1/#2}\textcolor{red!#3}{\rule{\BLACKBARSIZE\BARSIZE}{1.6ex}}%
}

\newcommand*\percent[5][v]{%
\ifx q#1%
    \np{#2}/\np{#3}(\ShowPercentage{#2}{#3})\else%
\ifx w#1%
    \np{#2}(\ShowPercentage{#2}{#3})\else%
\ifx m#1%
    \np{#2}%
    \inlinechart{#2}{#3}\else%
\ifx t#1%
    \FPprint{#2}%
    \hspace*{0.5ex}%
    \inlinechart{#2}{#3}\else%
\ifx p#1%
    \FPprint{#4}%
    \hspace*{0.5ex}%
    \pinlinechart{#2}{#3}{#5}\else%
\ifx n#1%
    \FPprint{#4}%
    \hspace*{0.5ex}%
    \ninlinechart{#2}{#3}{#5}\else%
\ifx c#1%
    \inlinechart{#2}{#3}{#5}%
\else%
    \np{#2}%
    \ifx r#1%
        /\np{#3}%
    \fi%
    \hspace*{0.5ex}(\ShowPercentage{#2}{#3}) %
    \inlinechart{#2}{#3}%
    \xspace
\fi\fi\fi\fi\fi%
}

\ifCLASSINFOpdf
\else
\fi
\definecolor{gre}{RGB}{101, 191, 127}
\definecolor{gree}{RGB}{7, 135, 44}

\newtcolorbox{keytheory}{colback=green!10,enhanced,title=Theory of Model Usefulness for Tuning,
	attach boxed title to top left={xshift=0mm},boxrule=0pt,after skip=1cm,before skip=1cm,right skip=0cm,breakable,fonttitle=\bfseries,toprule=0pt,bottomrule=0pt,rightrule=0pt,leftrule=4pt,arc=0mm,skin=enhancedlast jigsaw,sharp corners,colframe=gree,colbacktitle=gre,boxed title style={
		frame code={ 
			\fill[gre](frame.south west)--(frame.north west)--(frame.north east)--([xshift=3mm]frame.east)--(frame.south east)--cycle;
			\draw[line width=1mm,gre]([xshift=2mm]frame.north east)--([xshift=5mm]frame.east)--([xshift=2mm]frame.south east);
			
			\draw[line width=1mm,gre]([xshift=5mm]frame.north east)--([xshift=8mm]frame.east)--([xshift=5mm]frame.south east);
		}
	}
}

\makeatletter
\tcbset{
    myhbox/.style 2 args={%
        enhanced, 
        breakable,
        colback=white,
        colframe=black!60,
        attach boxed title to top left={yshift*=-\tcboxedtitleheight}, 
        title={#2},
        boxed title size=title,
        boxed title style={%
            sharp corners, 
            rounded corners=northwest, 
            colback=tcbcolframe, 
            boxrule=0pt,
        },
        underlay boxed title={%
            \path[fill=tcbcolframe] (title.south west)--(title.south east) 
                to[out=0, in=180] ([xshift=5mm]title.east)--
                (title.center-|frame.east)
                [rounded corners=\kvtcb@arc] |- 
                (frame.north) -| cycle; 
        },
        #1
    }
}   
\makeatother

\newtcolorbox{myhbox}[2][]{%
    myhbox={#1}{#2}
}
\definecolor{mycolor}{rgb}{0.122, 0.435, 0.698}

\newcommand{\rev}[1]{\textcolor{black}{#1}}

\newtcbox{\mytag}{nobeforeafter,colframe=mycolor,colback=mycolor!30!white,boxrule=0.7pt,arc=0pt,
 boxsep=-3pt,left=6pt,right=6pt,top=4pt,bottom=5pt,tcbox raise base}
\renewcommand{\mytag}[1]{}

\newcommand{\approach}{\texttt{L2RAC}} %

\mathchardef\mhyphen="2D

\DeclareMathAlphabet\mathbfcal{OMS}{cmsy}{b}{n}

\newcolumntype{P}[1]{>{\centering\arraybackslash}m{#1}}
\newcolumntype{Y}{>{\centering\arraybackslash}X}

   \usepackage[utf8]{inputenc}
   \usepackage[T1]{fontenc}
   \newcommand*\quotefont{\fontfamily{LinuxLibertineT-LF}} 

\newcommand*\quotesize{30} 
\newcommand*{\openquote}
   {\tikz[remember picture,overlay,xshift=-3ex,yshift=-0.5ex]
   \node (OQ) {\quotefont\fontsize{\quotesize}{\quotesize}\selectfont``};\kern0pt}

\newcommand*{\closequote}[1]
  {\tikz[remember picture,overlay,xshift=3ex,yshift={#1}]
   \node (CQ) {\quotefont\fontsize{\quotesize}{\quotesize}\selectfont''};}

\colorlet{shadecolor}{white}

\newcommand*\shadedauthorformat{\emph} 

\newcommand*\authoralign[1]{%
  \if#1l
    \def\authorfill{}\def\quotefill{\hfill}
  \else
    \if#1r
      \def\authorfill{\hfill}\def\quotefill{}
    \else
      \if#1c
        \gdef\authorfill{\hfill}\def\quotefill{\hfill}
      \else\typeout{Invalid option}
      \fi
    \fi
  \fi}
{\authoralign{#1}
\ifblank{#2}
   {\def\shadequoteauthor{}\def\yshift{-2ex}\def\quotefill{\hfill}}
   {\def\shadequoteauthor{\par\authorfill\shadedauthorformat{#2}}\def\yshift{1ex}}
\begin{snugshade}\begin{quote}\openquote}
{\shadequoteauthor\quotefill\closequote{\yshift}\end{quote}\end{snugshade}}

\begin{document}
\bstctlcite{MyBSTcontrol}

	%
	
	


     \title{Specializing Meta Multi-Objectivization as Baseline for Software Engineering Optimization}
	
	
	
	
	\author{
		Chengjie~Chuan,~\IEEEmembership{}
		Pengzhou~Chen,~\IEEEmembership{}
		Tao Chen~\IEEEmembership{} %
        \IEEEcompsocitemizethanks{
		    \IEEEcompsocthanksitem Corresponding author: Tao Chen (email: t.chen@bham.ac.uk).
			\IEEEcompsocthanksitem Chengjie Chuan and Pengzhou Chen are with the School of Computer Science and Engineering, University of Electronic Science and Technology of China, China, 610056.
			\IEEEcompsocthanksitem Tao Chen is with the IDEAS Lab, School of Computer Science, University of Birmingham, UK, B15 2TT.
            \IEEEcompsocthanksitem Chengjie Chuan and Pengzhou Chen are also supervised in the IDEAS Lab.
		}
		\thanks{}
  }

	
	\IEEEtitleabstractindextext{%

\begin{abstract}

Given the discrete nature of many software engineering (SE) optimization problems, trapping at local optima remains a crucial issue in the optimizer designs. Meta Multi-Objectivization  (MMO) is an optimization model for overcoming local optima by introducing an extra, auxiliary objective. Yet, MMO is originally designed only for configuration tuning, leaving its suitability for other SE problems in doubt, particularly regarding the choice of the auxiliary objective. In this paper, we study the possibility and resolution for specializing MMO as a general baseline across a range of SE optimization problems, making it a fundamental counterpart for future SE optimizers that seek to mitigate the issue of local optima. We do so by providing both theoretical remit and empirical evidence: we analytically show how the choice of the auxiliary objective is crucial while demonstrating the effectiveness of MMO under \mytag{R3C1}\rev{nine SE optimization problems comprising 212 cases} when the right one can be chosen, i.e., MMO is able to outperform the other state-of-the-art and general-purpose optimizers in \mytag{R3C1}\rev{79\% and 94\%} of the cases, respectively. Drawing on the theoretical understanding and empirical findings, we then propose \approach, a tool that automatically recommends the best auxiliary objective choice for a given SE problem/case, hence helping to specialize MMO therein. Through extensive experiments, we reveal that, in general, \approach~can more accurately rank the choice of auxiliary objective for MMO under a case than other guesses by up to 60\%, saving as much as 99\% of the cost that would otherwise be required to conduct trial-and-error when using MMO.

\end{abstract}


    }
	

	\maketitle

	\IEEEdisplaynontitleabstractindextext

	%
	\IEEEpeerreviewmaketitle

\section{Introduction}
\label{sec:introduction}


\IEEEPARstart{M}{any} software engineering (SE) tasks can be formulated as optimization problems, a.k.a. search-based software engineering (SBSE), including requirements planning~\cite{DBLP:journals/tse/XuanJRL12,DBLP:journals/ese/DurilloZAHN11}, software product line testing~\cite{DBLP:journals/tosem/XiangHLLLY24,DBLP:journals/tse/XiangHLLY22}, and  configuration tuning~\cite{DBLP:conf/ase/Ye0L26,DBLP:journals/tse/Nair0MSA20,DBLP:journals/tse/ChenCL24,DBLP:journals/jss/LiBHW25,chen2024adapting,DBLP:conf/ijcai/ChenCDS4RAG}. All those problems are either expensive and/or suffer from an intractable space, making manual or exhaustive optimization unrealistic. For example, it can take up to 166 minutes for measuring the performance of a single configuration~\cite{chen2024adapting}; the number of possible sequences/plans of fulfilling the requirements can be as large as $2^{1000}$~\cite{DBLP:journals/access/GengYJZLGX18}
As a result, over the past decades, many heuristic optimizers that search within the defined space for optimizing an objective, have been proposed in the SE community~\cite{DBLP:journals/access/GengYJZLGX18,DBLP:journals/tsc/XieWSSW23,DBLP:journals/tosem/XiangHLLLY24,DBLP:conf/etfa/FischerT12,DBLP:journals/tse/GouesNFW12,DBLP:journals/tosem/MkaouerKSKBD015,DBLP:journals/tse/TawosiSPH22}.

Albeit their remarkable achievements, existing optimizers for SE optimization still suffer from a long-standing issue, namely \textbf{stagnation at local optima}, in which the search is trapped at locally optimal but globally suboptimal solutions.  
Many prior studies have revealed the prevalence and severity of local optima issues in SE optimization, due primarily to the discrete nature. For example, in software configuration tuning, the local optima issue is primarily due to the way that configuration option is designed (e.g., a single digit change may replace the entire module), leading to high sparsity and ruggedness in the configuration landscape~\cite{DaL,DBLP:journals/pacmse/Gong024,DBLP:journals/tse/GongCB25,DBLP:conf/icse/XiangChen26}, which is known to be challenging for optimizers~\cite{DBLP:conf/kbse/XiongC25,DBLP:journals/tse/Nair0MSA20,DBLP:journals/corr/abs-2507-05995,DBLP:journals/tse/ChenCL24,DBLP:conf/icse/Ye0L25}. 
Yet another example is the software next release problem, in which the scale of requirements and dependency constraints creates a complex landscape, where the optimizers can be easily misled by the large basins of suboptimal solutions~\cite{DBLP:journals/tse/XuanJRL12,durillo2009study,DBLP:journals/infsof/BagnallRW01,DBLP:journals/ese/DurilloZAHN11}.

\definecolor{BlueCell}{RGB}{210,230,255}
\begin{table*}[t!]
\centering
\footnotesize
\caption{Exampled work that mitigates local optima from the SE community between 2008 and 2024.}
\label{tb:sbse_local_optima_strategies}

\begin{adjustbox}{width=\textwidth,center}
\begin{tabular}{llllll}
\toprule
\textbf{Optimizer} & \textbf{SE Optimization Problem} & \textbf{Local Optima Mitigation} & \textbf{Venue} & \textbf{Year} & \textbf{Reference}  \\ 
\midrule

\rowcolor{BlueCell}\texttt{PromiseTune} 
& Software Configuration Tuning (SCT) 
& Balance exploitation/exploration via causal rules 
& ICSE & 2026 & \cite{DBLP:journals/corr/abs-2507-05995} \\

\texttt{MAP-Elites} 
& Software Product Lines Testing (SPLT) 
& Improve solutions diversity using grid-based partitioning 
& TOSEM & 2024 & \cite{DBLP:journals/tosem/XiangHLLLY24} \\

\rowcolor{BlueCell}\texttt{NS} 
& Software Product Lines Testing (SPLT) 
& Improve solutions diversity using novelty search 
& TSE & 2022 & \cite{DBLP:journals/tse/XiangHLLY22} \\

\texttt{AutoCCAG} 
& Software Product Lines Testing (SPLT) 
& Improve solution diversity via complementary solutions discovery 
& ICSE & 2021 & \cite{DBLP:conf/icse/LuoLCCHQZLZWRZ21} \\

\rowcolor{BlueCell}\texttt{SATune} 
& Software Configuration Tuning (SCT) 
& Probabilistically accept inferior solutions
& ASE & 2021 & \cite{DBLP:conf/kbse/KocMWFP21} \\

\texttt{CONEX} 
& Software Configuration Tuning (SCT) 
& Probabilistically accept inferior solutions 
& TSE & 2020 & \cite{krishna2020conex} \\

\rowcolor{BlueCell}\texttt{FLASH} 
& Software Configuration Tuning (SCT) 
& Handle landscape sparsity via surrogate model
& TSE & 2020 & \cite{DBLP:journals/tse/Nair0MSA20} \\

\texttt{ROME} 
& Software Effort Estimation (SEE) 
& Handle landscape sparsity via surrogate model 
& TSE & 2020 & \cite{xia2020sequential} \\

\rowcolor{BlueCell}\texttt{FastCA} 
& Software Product Lines Testing (SPLT) 
& Enforce exploration by random adjustment 
& FSE & 2019 & \cite{DBLP:conf/sigsoft/LinCLLZ19} \\

\texttt{HGA} 
& Next Release Problem (NRP) 
& Enforce exploration by adapting operator choices  
& TOSEM & 2018 & \cite{zhang2018empirical} \\


\rowcolor{BlueCell}\texttt{ACO-L} 
& Software Project Scheduling (SPS) 
& Random perturbation based on identified local optima 
& TSE & 2012 & \cite{chen2012ant} \\

\texttt{BMA} 
& Next Release Problem (NRP) 
& Accept inferior solutions with random restarts 
& TSE & 2012 & \cite{DBLP:journals/tse/XuanJRL12} \\

\rowcolor{BlueCell}\texttt{CCIT} 
& Software Product Lines Testing (SPLT) 
& Prune search space via constrained local optima 
& TSE & 2008 & \cite{cohen2008constructing} \\

\bottomrule
\end{tabular}
\end{adjustbox}
\end{table*}

To mitigate local optima, the SE community has explored various mitigation mechanisms within the optimizers, as shown in Table~\ref{tb:sbse_local_optima_strategies}. Despite a community-wide effort to tackle them, overcoming local optima remains a ``tough nut to crack'' for many SE optimization problems; the current optimizers are still far from being effective general baselines for overcoming local optima across different SE domains.

Many prior designs of local optima mitigation in SE optimizers, as shown from Table~\ref{tb:sbse_local_optima_strategies}, have focused on the underlying algorithms. As an alternative, the recently proposed meta multi-objectivization (MMO)~\cite{chen2024adapting,DBLP:journals/tse/ChenCL24,DBLP:conf/sigsoft/0001L21} \mytag{R2C1}\rev{tackles} this challenge from a completely different perspective: instead of focusing on the algorithm, it seeks to overcome difficult local optima by seamlessly transforming the objective space, with no or limited changes required on the underlying algorithm. In essence, the core idea of MMO is to introduce an auxiliary objective (whose value we do not care about), which is optimized alongside the target objective in a specific (yet simple) way. Here, some extents of incomparability can be created in the newly-formed bi-objective space, hence promoting better diversity that helps to jump out from the local optima~\cite{DBLP:journals/tcyb/MaHLQWZ23}. For example, in configuration tuning, it has been shown that using MMO to optimize latency (target) and CPU load (auxiliary) leads to better latency results than optimizing latency alone~\cite{DBLP:journals/tse/ChenCL24}.

MMO is a theoretically fruitful yet simplistic method since it merely converts the objective space, which can be paired with any (Pareto) search algorithm. This makes it a promising candidate as a general baseline for overcoming local optima in SE optimization for two reasons: (1) it is general at the optimization-model level rather than tied to a specific search algorithm, and (2) it can be specialized for a particular SE problem and case by changing the auxiliary objective. These properties allow MMO to serve as a simple, configurable, and challenging reference point for evaluating future SE optimizers designed to overcome local optima. However, despite the emerging benefit of MMO, it has two major unaddressed issues:

\begin{itemize}
    \item MMO requires one to choose an auxiliary objective, which is important for the effectiveness, depending on how it correlates with the target objective. However, the choice and form of the auxiliary objective, as well as how it might influence the outcome, remain unclear.
    \item MMO is initially proposed for the configuration tuning problem. As such, while the notion of MMO is general and its simplicity is domain-transferrable, whether it can be effective for other SE optimization problems is still unknown, as they may exhibit characteristics that differ from those of configuration tuning, e.g., the correlation between target and auxiliary objectives~\cite{DBLP:journals/tosem/ChenL23,DBLP:conf/ssbse/MkaouerKBC14}.
\end{itemize}

To bridge the above gaps, in this paper, we contribute to the field of general SE optimization by \mytag{R2C1}\rev{promoting} MMO as a ``\textit{\textbf{simplistic baseline}}'' for overcoming local optima through ``\textit{\textbf{specialization}}''---automatically recommending the right auxiliary objective for the target SE optimization problem/case. Notably, we make contributions in several aspects:

\begin{itemize}
    \item \textbf{The theory and choice of auxiliary objective in MMO:} We provide an in-depth theoretical analysis on why and how the choice of the auxiliary objective is important to MMO. To understand the possible choices, we also turn to the literature of general multi-objectivization and identify seven fundamental forms of auxiliary objective choices.

    \item \textbf{An empirical study of MMO with different choices of auxiliary objective for diverse SE optimization problems:} We conduct a large-scale empirical study to evaluate how well MMO \mytag{R2C1}\rev{works} across the identified forms of auxiliary objectives. The study covers \mytag{R3C1}\rev{212 cases from nine SE optimization problems}. 

    \item \textbf{Automatically recommending the auxiliary objective of MMO for SE optimization problems:} Drawing on the findings, we propose a predictive tool, dubbed \approach, to automatically recommend suitable auxiliary-objective choices, given the characteristics of a case of an SE optimization problem~\cite{liefooghe2019landscape, DBLP:conf/gecco/LiefoogheVLZM21, DBLP:conf/gecco/MersmannBTPWR11,pearson1895contributions,decarlo1997meaning}.
\end{itemize}


In particular, we have revealed that:

\begin{itemize}
    \item MMO with the best form of auxiliary objective outperforms the original MMO and another multi-objectivization approach in \mytag{R3C1}\rev{93\%} of the cases;
    \item it also exceeds the state-of-the-art optimizers for individual SE optimization problems in \mytag{R3C1}\rev{79\%} of the cases;
    \item \mytag{R2C1}\rev{it performs} significantly better than other general-purpose optimizers for SE optimization problems, such as \texttt{SWAY}~\cite{DBLP:journals/tse/ChenNKM19}, in \mytag{R3C1}\rev{94\%} of the cases.
    \item \approach~also recommends promising choices of auxiliary objective that are considerably better than a random-guessing baseline and the generally best choices from empirical study, with up to 60\% improvement on NDCG, saving a significant amount of resources that would otherwise be required for trial-and-error.
\end{itemize}

With the support from \approach, what we advocate is that MMO can serve as a simplistic, easy-to-apply/compare yet challenging baseline for SE optimization problems, especially for newly proposed optimizers that aim to overcome the issue of local optima exhibited in an SE task. All code, datasets, and experimental results are publicly available at: \textcolor{blue}{\texttt{\href{https://github.com/ideas-labo/mmo-sbse}{https://github.com/ideas-labo/mmo-sbse}}}.

The remainder of this paper is organized as follows. Section~\ref{sec:background} introduces background and motivation. Section~\ref{Sec: Analysis} presents the theoretical analysis of the auxiliary objective choice for MMO. Section~\ref{sec:aux} specifies the common choices of auxiliary objectives. Section~\ref{sec:methodology} details our methodology. Section~\ref{sec:results} discusses the empirical results. 
Section~\ref{sec:tool} delineates our proposed predictive tool \approach. Section~\ref{sec:threats} analyzes threats to validity. Section~\ref{sec:related} discusses the related work while Section~\ref{sec:conclusion} concludes the paper.
\mytag{R3C5}\rev{The appendices provide in-depth details on the auxiliary-objective definitions, SE problems, and landscape features used in this work.}

\section{Preliminaries}
\label{sec:background}


\subsection{SE Optimization}

In SBSE, SE tasks can be formulated as optimization problems solved by using search algorithms~\cite{harman2001search,harman2012search}. The core principle of this paradigm is to automatically explore a large, complex, and often discrete space of possible solutions by optimizing an objective function that quantifies solution quality. Formally, let $S$ denote the search space, representing all possible candidate solutions for a given SE task. A solution $s \in S$ may correspond to a test suite, a software design, or a project plan. Let $f: S \rightarrow \mathbb{R}$ be an objective function that assigns a real-valued score to each candidate solution $s$. The objective is to find the optimal solution $s^*$ that minimizes/maximize this objective function\footnote{In this work, we focus on the single objective case, as multiple conflicting objectives can be aggregated via preferred weights~\cite{DBLP:journals/cor/XiaoAT13,DBLP:journals/tse/TawosiSPH22,DBLP:journals/apin/MafarjaTATADT23,DBLP:journals/cj/JuDH23}.}:
\begin{equation}
s^{*} = \arg\min_{s \in S} f(s) \text{ or } \arg\max_{s \in S} f(s)
\end{equation}
The objective can vary, ranging from coverage of code to the runtime of the software. The virtual and measurable nature of software artifacts, combined with software metrics, makes this domain particularly amenable to search algorithms~\cite{harman2004metrics,harman2012search}.


\subsection{Multi-Objectivization}

\mytag{R2C4}\rev{Multi-objectivization is different from the general problem of multi-objective optimization in the following:}

\begin{itemize}
    \item \mytag{R2C4}\rev{In standard multi-objective optimization, the problem naturally contains multiple objectives of interest, and the goal is often to approximate a Pareto front that represents trade-offs among these objectives. In contrast, multi-objectivization still targets a single-objective optimization problem, where only one objective is of final interest; the auxiliary objective is introduced to reshape the search space and mitigate local optima.}
    \item \mytag{R2C4}\rev{Multi-objective optimization produces a solution set, whereas multi-objectivization ultimately selects a single solution according to the target objective.}
\end{itemize}

As an important optimization paradigm, the core idea of multi-objectivization is to alter the characteristics of the search space by increasing the objective dimension, promoting the ability to escape from local optima traps~\cite{DBLP:conf/emo/KnowlesWC01,DBLP:journals/eor/Garza-FabrePR15,lochtefeld2015multi,DBLP:conf/ssbse/BuzdalovaBP13}. In SE optimization, two commonly adapted multi-objectivization techniques are  Plain Multi-Objectivization (PMO) ~\cite{DBLP:conf/kbse/DerakhshanfarDZ20,DBLP:conf/ssbse/MkaouerKBC14,DBLP:journals/ese/MkaouerKCHD17} and MMO~\cite{chen2024adapting,DBLP:journals/tse/ChenCL24,DBLP:conf/sigsoft/0001L21}. 



\subsubsection{Plain Multi-Objectivization Model (PMO)}

PMO is a naive and the simplest way of multi-objectivizing the problem by adding an auxiliary objective to be optimized simultaneously with the target one. Formally, it is expressed as\footnote{For simplicity, we assume minimization; maximizing objective can be converted via linear inversion.}:
\begin{equation}
\text{minimize} \begin{cases}
f_t(\bm{x}) \\
f_a(\bm{x})
\end{cases}
\label{eq:pmo}
\end{equation}
whereby \( f_t(\bm{x}) \) and \( f_a(\bm{x}) \) denote the target and auxiliary objective, respectively. Solutions in the search space are compared according to Pareto dominance relations\footnote{\mytag{R3C5}\rev{A solution $A$ Pareto-dominates $B$ if it is no worse than $B$ in all objectives and better in at least one objective; otherwise, $A$ and $B$ are non-dominated with respect to each other.}}. This partial order allows solutions that are similar in performance in the original single-objective space to become incomparable, thereby helping to maintain diversity and preventing premature convergence of the search process. PMO has been widely applied to several SE optimization cases, e.g., crash reproduction \cite{DBLP:conf/kbse/DerakhshanfarDZ20}, code refactoring \cite{DBLP:conf/ssbse/MkaouerKBC14,DBLP:journals/ese/MkaouerKCHD17}, and test generation \cite{DBLP:conf/ssbse/BuzdalovaBP13}. 


Yet, the PMO has an inherent limitation: it requires simultaneous optimization of both the target and auxiliary objectives\cite{DBLP:journals/tse/ChenCL24,DBLP:conf/sigsoft/0001L21}, which is unnecessary since we do not care about the values of the auxiliary objective. This means the search process may allocate valuable budgets to solutions that perform well on the auxiliary objective but poorly on the target objective.


\subsubsection{Meta Multi-Objectivization Model (MMO)}


\mytag{R2C4}\rev{MMO improves PMO by addressing its issue of wasting computational budget to optimize the auxiliary objective, which is of no interest for the single-objective optimization problem. This is achieved by searching in a transformed MMO space.} Specifically, MMO~\cite{chen2024adapting,DBLP:journals/tse/ChenCL24,DBLP:conf/sigsoft/0001L21} can be expressed as: 
\begin{equation}
\text{minimize} \left\{ \begin{array}{l}
g_1(\bm{x}) = f_t(\bm{x}) + w \cdot f_a(\bm{x}) \\
g_2(\bm{x}) = f_t(\bm{x}) - w \cdot f_a(\bm{x})
\end{array} \right.
\label{eq:mmo}
\end{equation}
Here, \( w \) is a weight parameter used to adjust the influence of the auxiliary objective\footnote{In this work, we set \( w = 1 \) without any adaptation for simplicity and effectiveness, as suggested previously ~\cite{DBLP:journals/tse/ChenCL24}.}. Mathematically, MMO scales
(stretches or shrinks) the solutions along the \( f_a(\bm{x}) \) axis by
a factor of $w$, after which it then rotates the scaled solutions by $45^{\circ}$ clockwise and then dilates them on both \( f_t(\bm{x}) \) and \( f_a(\bm{x}) \) by a factor of $\sqrt{2}$~\cite{DBLP:journals/tse/ChenCL24}. Such a geometric transformation forms the MMO space. In this way, several key features can be preserved by MMO~\cite{DBLP:journals/tse/ChenCL24}, such as we are still interested in optimizing the target objective, while do not care about the value of the auxiliary objective, but merely leveraging its varying values to increase the diversity for jumping out from the local optima (the key difference to PMO).

\subsection{Motivation}
\label{first_motivation}
MMO is originally designed for software configuration tuning, under the consideration of three key properties~\cite{DBLP:conf/sigsoft/0001L21}:

\begin{itemize}
    \item \textbf{Property 1:} The problem to be optimized is rugged and sparse, containing difficult local optima.
    \item \textbf{Property 2:} The optimization budget is valuable.
    \item \textbf{Property 3:} The correlation between the target and auxiliary objective is not strictly certain, i.e., good target objectives might correspond to very high or very low values of the auxiliary objective.
\end{itemize}

When using MMO for other SE optimization problems, \textbf{Property 1} and \textbf{Property 2} might still hold due to their generality. However, \textbf{Property 3} could become an obstacle. 

\begin{figure}[t!]
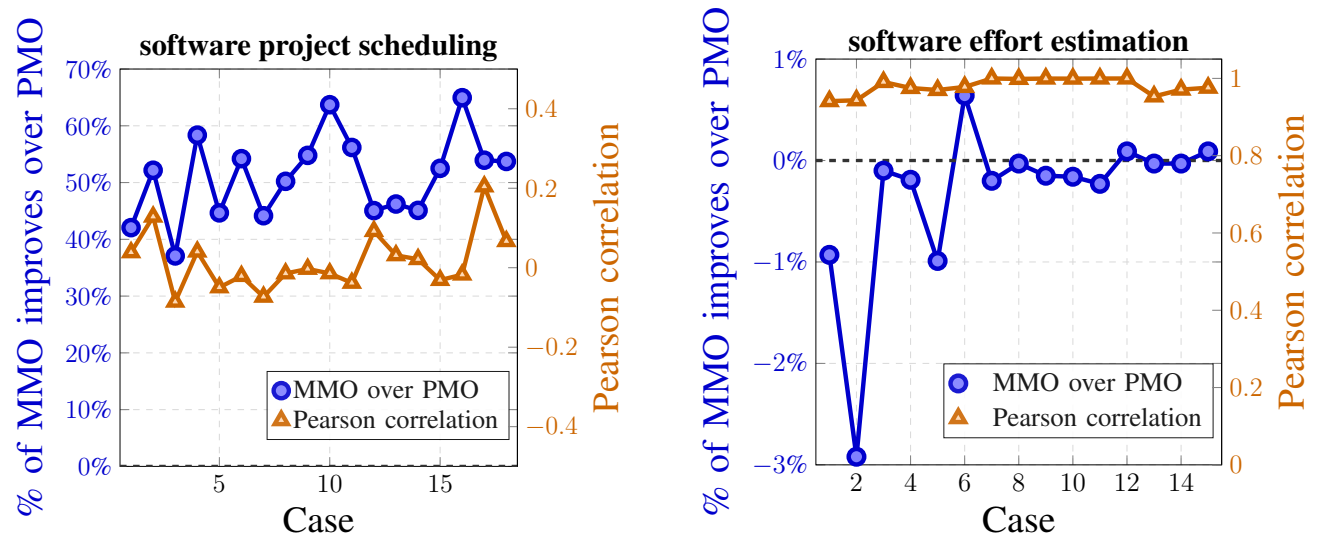

\centering
\subfloat[Less strictly certain]{
\includegraphics[width=.46\columnwidth]{Figures/weak_correlation.tex}
}
~
\subfloat[More strictly certain]{
\includegraphics[width=.46\columnwidth]{Figures/strong_correlation.tex}
}\\

\caption{Two SE optimization problems showing Pearson correlations between the target and the auxiliary objective (i.e., other properties of the software artifact), \% improvement of MMO over PMO where a negative value means PMO is better.}


\label{fig:correlation_case}
\end{figure}

In essence, \textbf{Property 3} implies that there should be neither strictly harmonic (linearly positively monotonic) nor conflicting (linearly negatively monotonic) relationships between target and auxiliary objectives; instead, they should exhibit uncertain correlation thereof. The original MMO design assumes that the auxiliary objective is chosen as another metric relevant to the software artifact to optimize, e.g., in software configuration tuning ~\cite{DBLP:conf/sigsoft/0001L21,DBLP:journals/tse/ChenCL24,DBLP:journals/tse/Nair0MSA20}, this can be CPU load, in addition to the target objective of latency. It has been shown that many software performance metrics, if not all, could fulfil \textbf{Property 3} ~\cite{DBLP:conf/icse/WeberKSAS23,DBLP:conf/oopsla/PintoCL14,DBLP:conf/sigsoft/0001L21}, but this may or may not be the case for other problems. In Figure~\ref{fig:correlation_case}, each case denotes a different instance of the corresponding SE optimization problem, with the same target and auxiliary objectives used within each subfigure; the figure illustrates how their correlation and the relative performance of MMO and PMO vary across problem instances. For example, Figure~\ref{fig:correlation_case}a shows the software project scheduling problem where the low Pearson values means that the correlation between the target and the chosen auxiliary objectives is rather uncertain, which aligns with \textbf{Property 3}, thus MMO constantly outperforms PMO. In contrast, optimizing software effort estimation in  Figure~\ref{fig:correlation_case}b shows a different case: the correlation thereof is clearly positive and monotonic in a rather linear fashion. This violates \textbf{Property 3}, hence MMO performs generally worse than PMO in most cases, despite its theoretical superiority.

The above motivate this work to investigate the choice/form of auxiliary objective for MMO, especially when it serves as a specialized baseline for other SE optimization problems.

\section{Theoretical Analysis of Auxiliary Objective} 
\label{Sec: Analysis}

\mytag{R2C5}\rev{Prior MMO work only offers brief discussions on \textbf{Property 3}~\cite{DBLP:journals/tse/ChenCL24,DBLP:conf/sigsoft/0001L21}. Combining the core logic of \textbf{Property 3} and the geometric transformation principles of MMO~\cite{DBLP:conf/sigsoft/0001L21}, here we conduct a theoretical analysis on why and how the auxiliary objective can influence MMO and its Pareto dominance relations in the MMO space.} \mytag{R3C5}\rev{Notably, applying MMO to a SE problem is simple; e.g., for software configuration tuning, $f_t$ can be runtime while $f_a$ can be a measured property of the system that is not the final target, e.g., CPU utilization.}


\mytag{R2C5}\rev{To provide the basis for the theoretical discussion of \textbf{Property 3}, it is crucial to understand how the choice of auxiliary objectives can impact the MMO ($w=1$), especially when it violates \textbf{Property 3}.} Considering the proposition:


\begin{displayquote}
\textit{\textbf{Proposition:} When $f_t$ and $f_a$ are linearly and monotonically correlated to each others (strictly harmonic or conflicting), e.g., $f_a=kf_t+\epsilon$, the solutions are either completely non-dominated or dominated in the MMO space.} 
\end{displayquote}


\textbf{Proof:} For any pair of solutions $x_i$ and $x_j$, the differences of meta-objectives in MMO are:
\begin{align*}
    g_{1}(x_i) - g_{1}(x_j) &= (f_{t}(x_i) - f_{t}(x_j)) + (f_{a}(x_i) - f_{a}(x_j)) \\
    g_{2}(x_i) - g_{2}(x_j) &= (f_{t}(x_i) - f_{t}(x_j)) - (f_{a}(x_i) - f_{a}(x_j))
\end{align*}
Since $f_t$ and $f_a$ have linearly monotonic correlation in the original space (assuming both are to be minimized, in the sense of PMO), i.e., $f_a=kf_t+\epsilon$, then we can have:
\begin{align*}
    g_{1}(x_i) - g_{1}(x_j) &= (1+k)(f_{t}(x_i) - f_{t}(x_j)) \\
    g_{2}(x_i) - g_{2}(x_j) &= (1-k)(f_{t}(x_i) - f_{t}(x_j)) 
\end{align*}
This leads to several mutually exclusive conditions for any pair of solutions, depending on the slope and correlation:
\begin{itemize}

\item If $f_t(x_i) < f_t(x_j) \implies f_a(x_i) < f_a(x_j)$ such that $x_i$ dominates $x_j$ in the original space, then $k>0$, hence when $0<k\leq1$, we have $g_1(x_i) < g_1(x_j)$ and $g_2(x_i) \leq g_2(x_j)$, meaning that $x_i$ dominates $x_j$ in the MMO space.

\item Same as above, when $k>1$, we have $g_1(x_i) < g_1(x_j)$ and $g_2(x_i) > g_2(x_j)$, meaning that $x_j$ and $x_i$ are non-dominated in the MMO space.

\item If $f_t(x_i) < f_t(x_j) \implies f_a(x_i) > f_a(x_j)$ such that $x_i$ and $x_j$ are non-dominated in the original space, then $k<0$, hence when $-1\leq k\leq0$, we have $g_1(x_i) \leq g_1(x_j)$ and $g_2(x_i) < g_2(x_j)$, meaning that $x_i$ dominates $x_j$ in the MMO space.

\item Same as above, when $k<-1$, we have $g_1(x_i) > g_1(x_j)$ and $g_2(x_i) < g_2(x_j)$, meaning that $x_j$ and $x_i$ are non-dominated in the MMO space.




\end{itemize}

\begin{figure}[t!]
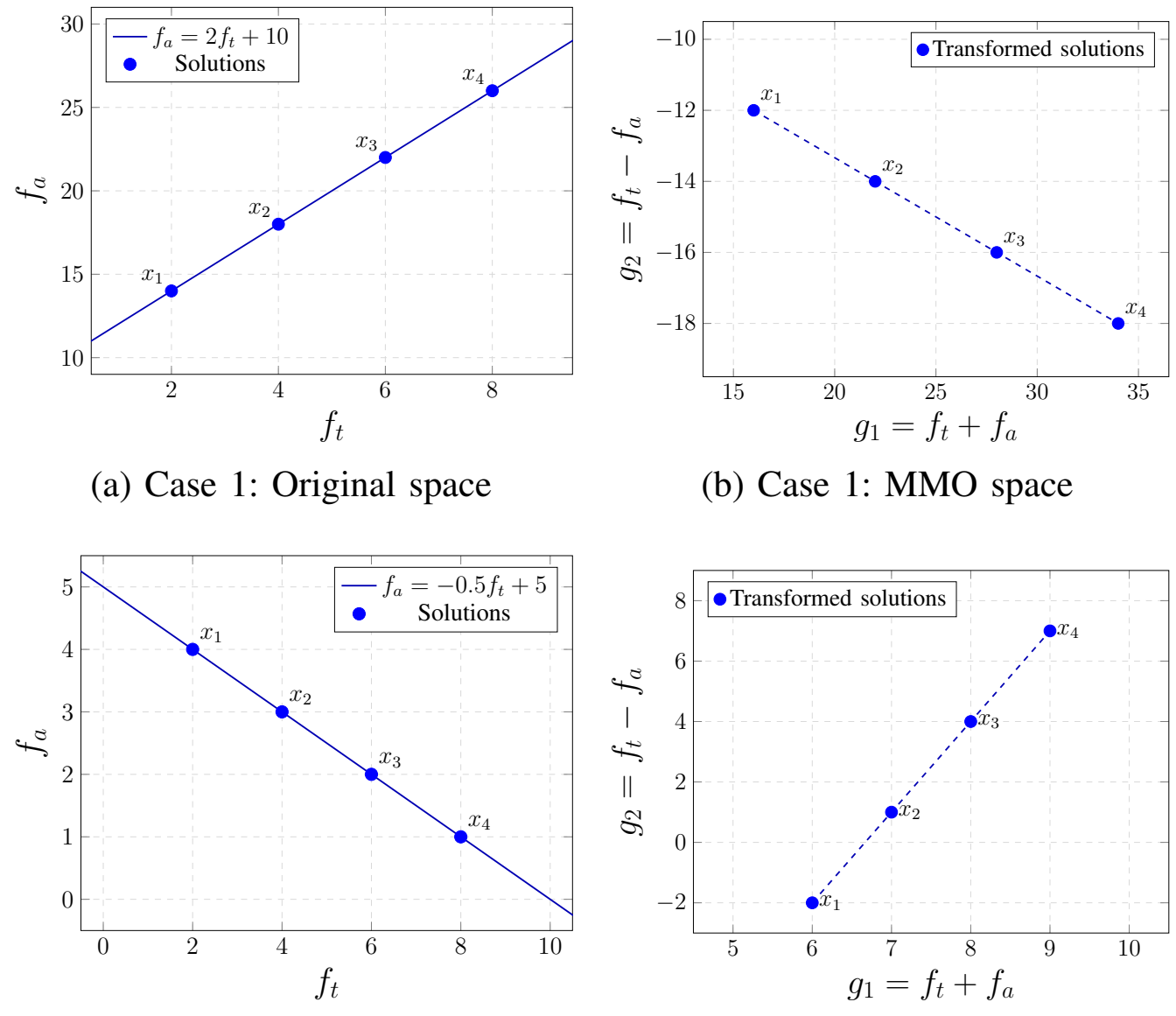

\centering
\subfloat[Case 1: Original space]{
\includegraphics[width=.235\textwidth]{Figures/theory1_before.tex} 
}
\subfloat[Case 1: MMO space]{
\includegraphics[width=.235\textwidth]{Figures/theory1_after.tex}
}\\

\subfloat[Case 2: Original space]{
\includegraphics[width=.235\textwidth]{Figures/theory2_before.tex}
}
\subfloat[Case 2: MMO space]{
\includegraphics[width=.235\textwidth]{Figures/theory2_after.tex}
}
\caption{The original and MMO space of two different cases that show the implication of auxiliary objective} 
\label{fig: MMO_case}
\end{figure}

Therefore, when $f_a$ is linearly monotonic to $f_t$, solutions in the MMO space can only be completely non-dominated or dominated by each other. Here we specify two examples:

\definecolor{BlueCell}{RGB}{210,230,255}
\begin{table*}[t!]
\centering
\caption{The generic choices/forms of auxiliary objective for multi-objectivization. The complexity is sorted by the mean runtime and its deviation required to calculate each auxiliary objective, the lower the level, the faster it can be computed.}
\label{tb:aux}
\begin{adjustbox}{width=\textwidth,center}
\begin{tabular}{
    l
    l
    l
    l
    l
}
\toprule
\textbf{Name} & \textbf{Formula} & \textbf{Description} & \textbf{Complexity} & \textbf{Reference} \\
\midrule

\rowcolor{BlueCell}
Reciprocal &
$f_a(x_i) = \frac{1}{\tilde{f}_t(x_i)}$ &
The reciprocal transformation to the normalized target objective of the solutions found so far. &Level 1 &
\cite{DBLP:conf/cec/WatanabeS05} \\

Gaussian Noise &
$f_a(x_i) = \tilde{f}_t(x_i) + \epsilon, \ \epsilon \sim \mathcal{N}(0,1)$ &
Target objective with a random and uncertain noise each evaluation. &Level 2 &
\cite{DBLP:conf/cec/WatanabeS05} \\

\rowcolor{BlueCell}
Age &
$f_a(x_i) = \text{age}(x_i)$ &
The age of explored solutions where solutions that are preserved longer are preferred more.&Level 3 &
\cite{DBLP:conf/emo/AbbassD03} \\


Original &
$f_a(x_i) = \text{property}(x_i)$ &
An extra property of the problem, e.g., CPU usage of a software in configuration tuning.&Level 4 &
\cite{DBLP:journals/ese/DurilloZAHN11,DBLP:journals/tpds/ZhuZLL16} \\

\rowcolor{BlueCell}
Penalty &
$f_a(x_i) = \tilde{f}_t(x_i) + \sum_{l \in L_i}\lambda_l p_l(x_i)$ &
Normalized target objective with a penalty based on the worse-performing solutions on local optima features. &Level 5 &
\cite{DBLP:journals/isci/Alsheddy18} \\

Grid Diversity &
$f_a(x_i) = D_{\mathrm{grid}}(x_i, P, \theta)$ &
Aggregate the target objective values of explored solutions in the grid space within a solution set $P$. &Level 6 &
\cite{DBLP:journals/tcyb/LiYL14} \\

\rowcolor{BlueCell}
Novelty &
$f_a(x_i) = \frac{1}{k} \sum_{j=1}^{k} d(x_i, x_j)$ &
The average distance in decisions space of neighboring solutions within a solution set. &Level 7 &
\cite{mouret2011novelty,lehman2011novelty} \\

\bottomrule
\end{tabular}
\end{adjustbox}
\end{table*}

\begin{itemize}

\item\mytag{R2C5}\rev{\textbf{Example 1:} As shown in Figure~\ref{fig: MMO_case}a, suppose that \(f_t\) and \(f_a\) have a linear, positive and monotonic correlation: $f_a = 2f_t + 10$}, \mytag{R2C5}\rev{where \(w=1\) in MMO and solution \(x_1=(2, 14)\) dominates solution \(x_2=(4, 18)\) in the original space. After the MMO transformation, we have \((g_1,g_2)(x_1)=(16,-12)\) and \((g_1,g_2)(x_2)=(22,-14)\). Thus, \(x_1\) is better on \(g_1\) but worse on \(g_2\), making the two solutions non-dominated in the MMO space (Figure~\ref{fig: MMO_case}b).}

\item\mytag{R2C5}\rev{\textbf{Example 2:} From Figures~\ref{fig: MMO_case}c and~\ref{fig: MMO_case}d, suppose that $f_t$ and $f_a$ have a linear, negative and monotonic correlation: $f_a = -0.5f_t + 5$}, \mytag{R2C5}\rev{where $w=1$. Suppose there are two non-dominated solutions $x_1 = (2, 4)$ and $x_2 = (4, 3)$ in the original space. After the MMO transformation, we have \((g_1,g_2)(x_1)=(6,-2)\) and \((g_1,g_2)(x_2)=(7,1)\). Hence, $x_1$ dominates $x_2$ in the MMO space.}

\end{itemize}

The above is a typical case where $f_t$ and $f_a$ largely violate \textbf{Property 3}, since they are linearly monotonic. Most importantly, the analysis suggests that an auxiliary objective, which linearly and monotonically correlates to the target, is certainly harmful to MMO---it has been shown that neither too many dominated nor too many non-dominated solutions in the MMO space is ideal~\cite{DBLP:journals/tse/ChenCL24}, since the former reduces MMO to the standard single-objective optimization (bias to exploitation) while the latter causes it to completely lose the search pressure for convergence (bias to exploration).

Indeed, in practice, $f_t$ and $f_a$ might not always exhibit a linearly monotonic correlation, but could be nonlinearly monotonic or non-monotonic; the above analysis only illustrates the extreme cases. Yet, importantly, such an analysis theoretically illustrates an important fact: the \(f_a\) directly determines its correlation to the \(f_t\), which also influences the extent of comparability between solutions in the MMO space, leading to direct impacts on MMO's effectiveness. Therefore, the choice of auxiliary objective needs to be able to maintain a correlation with the target such that a balanced proportion of dominated and non-dominated solutions can be preserved in the MMO space. This is exactly what is implied by \textbf{Property 3}.

Deriving from the analysis, given a set of solutions $\mathcal{S}$ in the original space of \(f_t\) and \(f_a\), we can infer several important features on the suitability of an MMO auxiliary objective when specializing for a SE optimization problem/case:

\begin{itemize}
    \item The slope of the $\mathcal{S}$ under \(f_t\) and \(f_a\).
    \item The correlation between \(f_t\) and \(f_a\) on $\mathcal{S}$.
    \item The extent of dominance within $\mathcal{S}$ against \(f_t\) and \(f_a\).
\end{itemize}

\section{The Choice and Forms of Auxiliary Objective}
\label{sec:aux}


To understand how auxiliary objectives can be constructed in MMO, we categorize them based on their origins and relationships with the target objective. Table~\ref{tb:aux} summarizes the details of seven most common auxiliary objectives used in multi-objectivization. Broadly, there are three categories:



\begin{itemize}
    \item \textbf{Problem-dependent objectives:} These objectives are directly derived from the optimization problem itself and naturally reflect its characteristics or requirements. Most of these objectives originate from multi-objective optimization research, e.g., the \textit{Original}, which uses an extra property of the problem as in the original MMO, and have also been used in some SBSE studies~\cite{DBLP:conf/kbse/DerakhshanfarDZ20,DBLP:conf/ssbse/MkaouerKBC14,DBLP:journals/ese/MkaouerKCHD17,DBLP:conf/ssbse/BuzdalovaBP13} to build multi-objectivization.

    \item \textbf{Solutions-derived objectives:} These objectives come from the properties of solutions explored previously rather than the problem itself, e.g., \textit{Age}, \textit{Novelty}, and \textit{Grid Diversity}, which promote exploration by favoring older, unique, or sparsely distributed individuals.

    \item \textbf{Transformed objectives:} These objectives are obtained by applying transformations or perturbations to the target objective, such as adding \textit{Gaussian Noise}, \textit{Penalty}, or \textit{Reciprocal}. Such transformations introduce nonlinear relationships with the target objective, relax dominance structures, and avoid excessively strong linear conflicts.
\end{itemize}


The above showcase a wide range of target--auxiliary relationships, providing additional design choices for adopting MMO to different SE optimization problems for fulfilling \textbf{Property 3}. Detailed discussions and formulas of each auxiliary objective choice/form can be found in Appendix A.

\begin{table*}[t!]
\centering
\scriptsize
\caption{Studied SE optimization problems and objective settings.}
\label{tb:sbse_instances}
\label{tb:rq1_objective_settings}
\definecolor{BlueCell}{RGB}{210,230,255}
\begin{adjustbox}{width=\textwidth,center}
\begin{tabular}{@{}lcllll@{}}
\toprule
\textbf{SE problem} & \textbf{$\#$ Cases} & \textbf{Search Space} & \rev{\textbf{Target objective}} & \rev{\textbf{\emph{Original} auxiliary objective}} & \textbf{Ref.} \\ 
\midrule

Next Release Problem (NRP) & 17 & $2^{100}\text{--}2^{1000}$ & \rev{Client profit} & \rev{Development cost} & \cite{DBLP:journals/tse/XuanJRL12} \\

\rowcolor{BlueCell}Software Configuration Tuning (SCT) & 26 & $864\text{--}3.32\times10^{13}$ & \rev{System property (e.g., runtime)} & \rev{Other property (e.g., CPU)} & \cite{DBLP:conf/sigsoft/0001L21,DBLP:journals/tse/ChenCL24} \\

Software Effort Estimation (SEE) & 15 & $2.76\times10^{132}\text{--}4.09\times10^{183}$ & \rev{Absolute effort error} & \rev{Estimation confidence interval} & \cite{DBLP:journals/tse/TawosiSPH22} \\

\rowcolor{BlueCell}Software Defect Prediction (SDP) & 40 & $2^{20}\text{--}2^{40}$ & \rev{AUC/accuracy} & \rev{Number of selected features} & \cite{DBLP:journals/jss/NiCWSG19} \\

Software Product Lines Testing (SPLT) & 21 & $1.1\times10^{38}\text{--}3.58\times10^{227}$ & \rev{$t$-wise product coverage} & \rev{Number of selected products} & \cite{DBLP:journals/tosem/XiangHLLLY24} \\

\rowcolor{BlueCell}Software Project Scheduling (SPS) & 18 & $10^{50}\text{--}4.96\times10^{664}$ & \rev{Makespan} & \rev{Project cost} & \cite{DBLP:conf/gecco/ChicanoLNA11} \\

Third Party Library Migration (TPLM) & 35 & $2^{30}\text{--}2^{460}$ & \rev{Migration similarity} & \rev{Number of migrations/replacements} & \cite{DBLP:conf/evoW/DeshpandeMOS22} \\

\rowcolor{BlueCell}Workflow Scheduling (WS) & 20 & $1.41\times10^{66}\text{--}1.07\times10^{4009}$ & \rev{Workflow makespan} & \rev{Execution cost} & \cite{DBLP:journals/tpds/ZhuZLL16} \\

Web Service Composition (WSC) & 20 & $1.2\times10^{61}\text{--}1.65\times10^{102}$ & \rev{QoS property (e.g., throughput)} & \rev{Other QoS property (e.g., latency)} & \cite{DBLP:conf/gecco/0001LY18,zheng2012investigating} \\

\bottomrule
\end{tabular}
\end{adjustbox}
\footnotesize{
\begin{flushleft}
The reported search spaces include infeasible solutions. For scenarios involving continuous decision variables, the search space is specified using 32-bit floating-point precision. The reported sizes are approximate and serve as indicative measures rather than exact values.
\end{flushleft}
}
\end{table*}

\section{Empirical Study Methodology}
\label{sec:methodology}


\subsection{Research Questions}

To facilitate our analysis, for each case of a SE optimization problem, we define two MMO variants: \textbf{MMO\textsubscript{b}}, which denotes the MMO equipped with the best auxiliary objective choice for such a case over repeated runs; and the original MMO design that \mytag{R2C1}\rev{uses} the extra property of the problem as the auxiliary objective~\cite{DBLP:journals/tse/ChenCL24}, denoted as \textbf{MMO\textsubscript{o}}. 

\mytag{R2C6}\rev{To identify MMO\textsubscript{b}, we first apply the non-parametric Scott--Knott Effect Size Difference (ESD) test~\cite{tantithamthavorn2017mvt} to group and rank the MMO results associated with different auxiliary-objective choices according to their target-objective performance. If multiple choices have the same rank, we further distinguish them by comparing their median target-objective values. If ties still remain, the final ranking is determined based on the predefined complexity levels from Table~\ref{tb:aux}, where lower-complexity auxiliary objectives are preferred. The top-ranked auxiliary-objective choice is used as MMO\textsubscript{b} for that case.}

\mytag{R2C8}\rev{In this study, \emph{Original} denotes the problem-native auxiliary objective adopted from the formulation or common setup of each SE problem; it does not represent an exhaustive set of all possible problem-native objectives. Likewise, MMO\textsubscript{b} denotes the best-performing choice within the seven auxiliary-objective forms studied in this work, rather than the globally best auxiliary objective among all possible formulations. For consistency with prior PMO applications~\cite{DBLP:conf/kbse/DerakhshanfarDZ20,DBLP:conf/ssbse/MkaouerKBC14,DBLP:journals/ese/MkaouerKCHD17,DBLP:conf/ssbse/BuzdalovaBP13}, PMO uses the same \emph{Original} auxiliary objective as MMO\textsubscript{o}. Consequently, our comparison evaluates MMO\textsubscript{b} against this established PMO setting; it does not compare MMO and PMO after independently optimizing the auxiliary-objective choice for each model.}

From the above, we study several research questions (RQs):



\keybox{
\textbf{RQ1:} How well can {MMO\textsubscript{b}} perform against PMO and {MMO\textsubscript{o}} for SE optimization?
}

\textbf{RQ1} is important to confirm the necessity of crossing a best auxiliary objective for each case. The next natural question is:

\keybox{
\textbf{RQ2:} What is the effectiveness of MMO\textsubscript{b} against the state-of-the-art methods in individual SE optimization problems?
}

Since we seek to position MMO as a general baseline for SE optimization, a crucial question is:

\keybox{
\textbf{RQ3:} How does MMO\textsubscript{b} perform against the other general-purpose algorithms for SE optimization?
}


\subsection{SE Optimization Problems}

To evaluate the generality of MMO, we selected nine representative SE optimization problems via the criteria below:

\begin{itemize}
    \item The problem must have publicly available datasets;
    \item To avoid the noises of solution representation, we focus on problems with tabular variables rather than texts/code.
    \item Each problem should include at least two measurable metrics as the objectives, which allows the original MMO to leverage characteristics inherent to the problem itself.
\end{itemize}

We use all the publicly available cases for the SE optimization problem, unless it exceeds the capability of our resources and does not complete one run of all experiments within, e.g., 24 hours. 
As from Table~\ref{tb:sbse_instances}, in this work, we study \mytag{R3C1}\rev{nine} SE optimization problems of diverse cases and complexities.

\mytag{R3C1}\rev{Indeed, certain SE optimization problems may not be typical, i.e., SCT. Yet, there are strong reasons for its inclusion:}
\begin{itemize}
    \item \mytag{R3C1}\rev{The original proposal of MMO specifically targeted SCT~\cite{DBLP:conf/sigsoft/0001L21,DBLP:journals/tse/ChenCL24}. This work goes beyond that by investigating the generality of MMO across a wide range of SE problems, and hence it remains important to verify that the conclusion does not change with respect to the SE problem where MMO was originally proposed.}
    \item \mytag{R3C1}\rev{There has been a strong and increasing effort on SCT from the SE perspective, since ensuring configuration performance is a crucial engineering task~\cite{conex,DBLP:journals/tse/Nair0MSA20,BOCA,DBLP:journals/tosem/ChenL23a}.}
\end{itemize}


It is worth noting that while some SE optimization problems studied involve multiple objectives, there are cases where a single objective is the most important, which serves as the target with another objective is used as the auxiliary objective by the original MMO. In this work, we select the target objective corresponding to the most important one, as stated in the literature. More details can be found in Appendix B.

\mytag{R3C3}\rev{Applying different auxiliary objectives $f_a$ in Table~\ref{tb:aux} to a case of an SE problem is straightforward, and some forms depend on the target objective $f_t$ (assuming both $f_t$ and $f_a$ are normalized when applicable):}

\begin{itemize}
    \item \mytag{R3C3}\rev{\textbf{Reciprocal:} This simply inverts the $f_t$ to be the $f_a$, which is agnostic to the SE problems. For example, in WSC, a solution uses throughput as $f_t$, then it has $f_a=\frac{1}{f_t}$.}
     \item \mytag{R3C3}\rev{\textbf{Gaussian Noise:} This means the $f_a$ consists of $f_t$ and some noise, and it is the same across the SE problems. For example, in SDP, a solution with accuracy as $f_t$ might have $f_a=f_t+0.56$, where $0.56$ is a random noise.}
    \item \mytag{R3C3}\rev{\textbf{Age:} Again, another auxiliary objective that is consistent over the SE problem. For example, in SPS, a solution with makespan as $f_t$ can have $f_a=5$, where $5$ indicates that the corresponding solution has survived for $5$ iterations.}
    \item \mytag{R3C3}\rev{\textbf{Original:} This is case/SE problem dependent, such that $f_a$ is another property of the software artifact to optimize, e.g., in SCT, $f_t$ is runtime while $f_a$ can be the CPU\%.}
    \item \mytag{R3C3}\rev{\textbf{Penalty:} Here, $f_a$ is a combination of $f_t$ and a penalty of the worst solution so far, selected based on local optima feature. For example, in SPS, a solution with project duration $f_t=0.40$ can have $f_a=0.50$, where the $0.10$ gap is the penalty of local optima-related features.}
    \item \mytag{R3C3}\rev{\textbf{Grid Diversity:} Here, $f_a$ is formed based on grid-based diversity measures of $f_t$, which is again SE problem agnostic. For example, in SPLT, the $f_t$ of a solution can be $t$-wise product coverage while $f_a=0.654$, where $0.654$ is the crowding/diversity distance of the solution within the grids of current population.}
    \item \mytag{R3C3}\rev{\textbf{Novelty:} Similar to the \textit{Grid Diversity}, this also uses $f_a$ as the diversity of a solution. For example, in NRP, a release plan may have $f_a=5$, which is the mean Hamming distance of the release plan to its neighboring release plans within the current population.}
\end{itemize}

\subsection{Compared Optimizers}


We use a set of existing optimizers for the SE optimization problems as the counterparts.





\begin{table}[t!]
\centering
\footnotesize
\caption{State-of-the-art single-objective optimizers for the SE optimization problems.}
\label{tb:sbse_single_baselines}

\begin{adjustbox}{width=\linewidth,center}
\begin{tabular}{llcl}
\toprule
\textbf{SE Optimization Problem} & \textbf{Optimizer} & \textbf{Year} & \textbf{Reference} \\
\midrule
Next Release Problem (NRP) 
& \texttt{BMA} & 2012 &  \cite{DBLP:journals/tse/XuanJRL12} \\

\rowcolor{BlueCell}Software Configuration Tuning (SCT) 
& \texttt{PromiseTune} & 2026 &  \cite{DBLP:journals/corr/abs-2507-05995} \\

Software Effort Estimation (SEE) 
& \texttt{GA} & 2022 &  \cite{DBLP:journals/tse/TawosiSPH22} \\

\rowcolor{BlueCell}Software Defect Prediction (SDP) 
& \texttt{SBWOA} & 2023 &  \cite{DBLP:journals/apin/MafarjaTATADT23} \\

Software Product Lines Testing (SPLT) 
& \texttt{LS-Sampling-Plus} & 2024 &  \cite{luo2024lssamplingplus} \\

\rowcolor{BlueCell}Software Project Scheduling (SPS) 
& \texttt{GP-ILS} & 2018 & \cite{DBLP:conf/gecco/SabarTS18} \\

Third Party Library Migration (TPLM) 
& \texttt{GA} & 2022 &  \cite{DBLP:conf/evoW/DeshpandeMOS22} \\

\rowcolor{BlueCell}Workflow Scheduling (WS) 
& \texttt{HGA} & 2016 &  \cite{DBLP:journals/jpdc/AhmadLMAK16} \\

Web Service Composition (WSC) 
& \texttt{HWOA} & 2023 &  \cite{DBLP:journals/cj/JuDH23} \\
\bottomrule
\end{tabular}
\end{adjustbox}
\vspace{0.5mm}
\begin{minipage}{\linewidth}
\footnotesize \mytag{R3C4}\rev{\textit{Note:} None of these optimizers uses multi-objectivization in our experiments.}
\end{minipage}
\end{table}

\subsubsection{State-of-the-art Optimizers for SE Optimization Problems}

Each SE optimization problem has a set of specifically designed optimizers, tailored to the unique problem characteristics. For example, \texttt{PromiseTune} ~\cite{DBLP:journals/corr/abs-2507-05995} is designed for handling the highly \mytag{R2C1}\rev{rugged/sparse} landscape of configurable systems. To that end, we select them \mytag{R2C1}\rev{accordingly} if they:

\begin{itemize}
    \item achieve the best performance compared with other \mytag{R2C1}\rev{optimizers} for the corresponding SE optimization problem;
    \item are published in \mytag{R2C1}\rev{major SE venues}, e.g., ICSE and TSE;
    \item are compatible with our experiment setup for fair comparisons. For example, those that terminate merely based on wall-clock time rather than the number of objective evaluations are ruled out, since the former can be highly sensitive to the hardware and programming language.
\end{itemize}

The optimizers are shown in Table~\ref{tb:sbse_single_baselines}. Note that some state-of-the-art optimizers assume a single, weighted-sum aggregation of multiple objectives, for which we simply optimize only the target objective without changing their heuristics.

\subsubsection{General-Purpose SE Optimizers}


Like MMO, there also exist various general optimizers from the SE community:

\begin{itemize}
    \item \textbf{\texttt{SWAY}~\cite{DBLP:journals/tse/ChenNKM19}} partitions the solution space and eliminate regions that are unlikely to contain high-quality solutions, tested on six SE optimization problems\cite{DBLP:journals/tse/ChenNKM19,DBLP:journals/tse/ChenGC25,lee2021preliminary,lustosa2024learning}.

      \item \textbf{\texttt{LINE}~\cite{DBLP:journals/corr/abs-2506-04509}} relies on diversity sampling, which selects representative solutions that cover structurally different regions of the search space. It has been used on five or more SE optimization problems.

    \item \textbf{\texttt{LITE}~\cite{DBLP:journals/corr/abs-2506-04509}:} This optimizer, being applied to at least five SE optimization problems, relies on label inference, which guides the search toward promising regions by learning from a small set of evaluated solutions. 
\end{itemize}

\label{sec:acc-metrics}

\begin{table}[t!]
\centering
\footnotesize
\caption{Budget and population size (only applied to population-based optimizers) for the SE problems.}
\label{tb:sbse_settings}
\definecolor{BlueCell}{RGB}{210,230,255}
\begin{adjustbox}{width=0.95\linewidth,center}
\begin{tabular}{llll}
\toprule
\textbf{SE Optimization Problem} & \textbf{Budget} & \textbf{Population Size} & \rev{\textbf{Reference}} \\
\midrule
Next Release Problem (NRP) & 100,000 & 100 & \rev{\cite{DBLP:journals/tse/XuanJRL12}} \\
\rowcolor{BlueCell}Software Configuration Tuning (SCT) & 400-2,500 & 20-60 & \rev{\cite{DBLP:conf/sigsoft/0001L21,DBLP:journals/tse/ChenCL24}} \\
Software Effort Estimation (SEE) & 25,000 & 100 & \rev{\cite{DBLP:journals/tse/TawosiSPH22}} \\
\rowcolor{BlueCell}Software Defect Prediction (SDP) & 3,000 & 100 & \rev{\cite{DBLP:journals/jss/NiCWSG19}} \\
Software Product Lines Testing (SPLT) & 12,000 & 100 & \rev{\cite{DBLP:journals/tosem/XiangHLLLY24}} \\
\rowcolor{BlueCell}Software Project Scheduling (SPS) & 100,000 & 100 & \rev{\cite{DBLP:conf/gecco/ChicanoLNA11}} \\
Third Party Library Migration (TPLM) & 25,000 & 100 & \rev{\cite{DBLP:conf/evoW/DeshpandeMOS22}} \\
\rowcolor{BlueCell}Workflow Scheduling (WS) & 50,000 & 50 & \rev{\cite{DBLP:journals/tpds/ZhuZLL16}} \\
Web Service Composition (WSC) & 5,000 & 100 & \rev{\cite{DBLP:conf/gecco/0001LY18,zheng2012investigating}} \\
\bottomrule
\end{tabular}
\end{adjustbox}
\end{table}




\subsection{Settings}

\mytag{R1C4}\rev{Since MMO is an optimization model rather than a low-level search algorithm, we pair it with NSGA-II~\cite{DBLP:journals/tec/DebAPM02}, as this has been shown to be the best in prior MMO studies~\cite{DBLP:conf/sigsoft/0001L21,DBLP:journals/tse/ChenCL24,chen2024adapting}. Further, NSGA-II is arguably the most common algorithm for SBSE and SE optimization studies with less than three objectives~\cite{durillo2009study,DBLP:conf/gecco/ChicanoLNA11,DBLP:conf/ssbse/MkaouerKBC14} (hence is also beneficial for PMO). As such, adopting NSGA-II allows us to compare them under the best conditions.} We adopt the partial duplicate retention mechanism in prior MMO studies~\cite{chen2024adapting}.  
The crossover and mutation operators (and parameters) follow the settings in prior work~\cite{DBLP:journals/tse/XuanJRL12,DBLP:journals/tse/ChenCL24,DBLP:journals/tse/TawosiSPH22,DBLP:journals/jss/NiCWSG19,DBLP:journals/tosem/XiangHLLLY24,DBLP:conf/gecco/ChicanoLNA11,DBLP:conf/evoW/DeshpandeMOS22,DBLP:journals/tpds/ZhuZLL16,DBLP:conf/gecco/0001LY18}.

For those optimizers that use population, we set a population size as used in their original studies. All the other hyperparameters are set as default or as what have been used by their authors. The termination criterion used in this work is the budget, measured by the number of evaluations to the objectives---a standard, hardware and implementation independent way that is widely used in SBSE ~\cite{DBLP:conf/sigsoft/0001L21,DBLP:journals/tse/ChenCL24,DBLP:journals/tse/Nair0MSA20}. \mytag{R3C5}\rev{The actual budget sizes are taken from the past work, because they have been shown to ensure good convergence for a wide range of optimizers while allowing us to maintain fair comparisons.} In particular, we set a maximum runtime of 24 hours to enable realistic completion of all experiments, such that the best solution found is returned even if the optimizers have not consumed all the budget within the 24-hour limit. The details have been presented in Table~\ref{tb:sbse_settings}.

For necessary constraint handling, we primarily adopt the same mechanisms as those used in their original studies~\cite{DBLP:journals/tpds/ZhuZLL16}.

\definecolor{BlueCell}{RGB}{210,230,255}

\begin{table*}[t!]
\centering
\scriptsize

\caption{Comparing  MMO\textsubscript{b} with MMO\textsubscript{o} and PMO across all cases and over 10 runs. The achieved target objective values are normalized. \colorbox{BlueCell}{Blue} denote the best optimizer for a SE optimization problem. Per case results can be found at: {\color{blue!60!black}\url{https://github.com/ideas-labo/mmo-sbse/blob/main/RQS/RQ1/RQ1.pdf}}.}

\label{tb:rq1_result}

\begin{adjustbox}{width=\linewidth,center}
\begin{tabular}{
l
>{\centering\arraybackslash}p{0.6cm} >{\centering\arraybackslash}p{0.6cm} >{\centering\arraybackslash}p{0.6cm}
>{\centering\arraybackslash}p{1.6cm} >{\centering\arraybackslash}p{1.6cm} >{\centering\arraybackslash}p{1.6cm}
>{\centering\arraybackslash}p{0.6cm} >{\centering\arraybackslash}p{0.6cm} >{\centering\arraybackslash}p{0.6cm}
>{\centering\arraybackslash}p{0.6cm} >{\centering\arraybackslash}p{0.6cm} >{\centering\arraybackslash}p{0.6cm}
}
\toprule
\multirow{2}{*}{\textbf{SE Optimization Problem}} &
\multicolumn{3}{c}{\textbf{Overall Rank}} &
\multicolumn{3}{c}{\textbf{Mean ± Std}} &
\multicolumn{3}{c}{\textbf{Ranked Best/Total}} &
\multicolumn{3}{c}{\textbf{Ranked Worst/Total}} \\

\cmidrule(lr){2-4} \cmidrule(lr){5-7} \cmidrule(lr){8-10} \cmidrule(lr){11-13}
& \textbf{MMO\textsubscript{b}} & \textbf{MMO\textsubscript{o}} & \textbf{PMO} 
& \textbf{MMO\textsubscript{b}} & \textbf{MMO\textsubscript{o}} & \textbf{PMO} 
& \textbf{MMO\textsubscript{b}} & \textbf{MMO\textsubscript{o}} & \textbf{PMO} 
& \textbf{MMO\textsubscript{b}} & \textbf{MMO\textsubscript{o}} & \textbf{PMO} \\
\midrule

Next Release Problem (NRP) &
\cellcolor{BlueCell}1.000 & \cellcolor{BlueCell}1.000 & 2.000 &
\cellcolor{BlueCell}$0.011 \pm 0.011$ &\cellcolor{BlueCell} $0.011 \pm 0.011$ & $0.703 \pm 0.152$ &
\cellcolor{BlueCell}17/17 & \cellcolor{BlueCell}17/17 & 0/17 &
\cellcolor{BlueCell}0/17 & \cellcolor{BlueCell}0/17 & 17/17 \\

Software Configuration Tuning (SCT) &
\cellcolor{BlueCell}1.150 & 1.770 & 1.850 &
\cellcolor{BlueCell}$0.199 \pm 0.140$ & $0.379 \pm 0.247$ & $0.390 \pm 0.207$ &
\cellcolor{BlueCell}22/26 & 12/26 & 9/26 &
\cellcolor{BlueCell}3/26 & 9/26 & 12/26 \\

Software Effort Estimation (SEE) &
\cellcolor{BlueCell}1.470 & 1.930 & \cellcolor{BlueCell}1.470 &
\cellcolor{BlueCell}$0.277 \pm 0.197$ & $0.411 \pm 0.226$ & $0.281 \pm 0.185$ &
\cellcolor{BlueCell}8/15 & 7/15 & 8/15 &
6/15 & 8/15 & \cellcolor{BlueCell}2/15 \\

Software Defect Prediction (SDP) &
\cellcolor{BlueCell}1.000 & 2.050 & 2.170 &
\cellcolor{BlueCell}$0.443 \pm 0.265$ & $0.555 \pm 0.269$ & $0.569 \pm 0.280$ &
\cellcolor{BlueCell}40/40 & 5/40 & 4/40 &
\cellcolor{BlueCell}0/40 & 24/40 & 29/40 \\

Software Product Lines Testing (SPLT) &
\cellcolor{BlueCell}1.000 & 1.380 & 2.330 &
\cellcolor{BlueCell}$0.242 \pm 0.135$ & $0.278 \pm 0.124$ & $0.655 \pm 0.192$ &
\cellcolor{BlueCell}21/21 & 13/21 & 1/21 &
\cellcolor{BlueCell}0/21 & \cellcolor{BlueCell}0/21 & 20/21 \\

Software Project Scheduling (SPS) &
\cellcolor{BlueCell}1.000 & 1.330 & 2.330 &
\cellcolor{BlueCell}$0.013 \pm 0.010$ & $0.015 \pm 0.011$ & $0.841 \pm 0.092$ &
\cellcolor{BlueCell}18/18 & 12/18 & 0/18 &
\cellcolor{BlueCell}0/18 & \cellcolor{BlueCell}0/18 & 18/18 \\

Third Party Library Migration (TPLM) &
\cellcolor{BlueCell}1.090 & 1.400 & 1.830 &
\cellcolor{BlueCell}$0.165 \pm 0.133$ & $0.201 \pm 0.130$ & $0.505 \pm 0.170$ &
\cellcolor{BlueCell}32/35 & 23/35 & 12/35 &
\cellcolor{BlueCell}1/35 & 6/35 & 23/35 \\

Workflow Scheduling (WS) &
\cellcolor{BlueCell}1.000 & 1.750 & 2.750 &
\cellcolor{BlueCell}$0.096 \pm 0.071$ & $0.130 \pm 0.086$ & $0.517 \pm 0.246$ &
\cellcolor{BlueCell}20/20 & 5/20 & 0/20 &
\cellcolor{BlueCell}0/20 & \cellcolor{BlueCell}0/20 & 20/20 \\

Web Service Composition (WSC) &
\cellcolor{BlueCell}1.000 & 1.550 & 2.400 &
\cellcolor{BlueCell}$0.337 \pm 0.194$ & $0.412 \pm 0.173$ & $0.665 \pm 0.209$ &
\cellcolor{BlueCell}20/20 & 10/20 & 0/20 &
\cellcolor{BlueCell}0/20 & 2/20 & 19/20 \\

\midrule[0.8pt]
\rev{\textbf{Average}} &
\cellcolor{BlueCell}\rev{\textbf{1.079}} & \rev{1.573} & \rev{2.126} &
\cellcolor{BlueCell}\rev{$\mathbf{0.198 \pm 0.128}$} & \rev{$0.266 \pm 0.142$} & \rev{$0.570 \pm 0.193$} &
\cellcolor{BlueCell}\rev{\textbf{198/212}} & \rev{104/212} & \rev{34/212} &
\cellcolor{BlueCell}\rev{\textbf{10/212}} & \rev{49/212} & \rev{160/212} \\
\bottomrule
\end{tabular}
\end{adjustbox}
\end{table*}

\subsection{\mytag{R3C2}\rev{Evaluation Metrics and Statistical Validation}}
\label{sec:sta}

\mytag{R3C2}\rev{Since many cases in the SE problem are intractable, we do not use the global optimum as the evaluation metric. Rather, we compare the best target objective values found in the evaluation. Since the objective values range differently across all cases for a SE problem, e.g., for SCT, one system can have runtime range between $100$ and $500$ seconds while another might be of $[0.5,10]$, we apply min--max normalization (where the max and min are chosen from the results returned by all methods for a case) to standardize the results across all cases for a SE problem, in which all objectives are converted to minimization, if needed. A normalization of 0.199 versus 0.75 means that the former achieves much better observed results than the latter for all cases of a SE problem in general (under the same budget), despite that it might not reach the global optimum. This is also a common practice in many SE optimization works with intractable problems~\cite{harman2001search,harman2012search,DBLP:journals/tse/ChenNKM19,DBLP:journals/tosem/ChenL23}.}

To mitigate stochastic bias, we use several statistical tests.

\subsubsection{Wilcoxon Signed Rank}



We use the Wilcoxon signed-rank test---a non-parametric statistical test widely used for pairwise comparisons ($\alpha = 0.05$)~\cite{DBLP:conf/icse/ArcuriB11}---to assess two methods over repeated runs on the same problem cases, where each run contains the same data samples.  



\subsubsection{Scott-Knott Effect Size Difference (ESD) Test}

When comparing multiple methods, we leverage the non-parametric Scott-Knott ESD test~\cite{tantithamthavorn2017mvt,DBLP:conf/icse/WangChen26}. In a nutshell, Scott-Knott ESD test sorts the list of treatments by their median values of the metric. Next, it splits the list into two sub-lists with the largest expected difference~\cite{xia2018hyperparameter}. For example, suppose that we compare $A$, $B$, and $C$, a possible split could be $\{A, B\}$, $\{C\}$, with the rank ($r$) of $1$ and $2$, respectively. This means that, in the statistical sense, $A$ and $B$ perform similarly, but they are significantly better than $C$. In contrast to other statistical tests on multiple comparisons (e.g., Kruskal-Wallis test~\cite{mckight2010kruskal}), Scott-Knott ESD test overcomes the confounding factor of overlapping groups without requiring post-hoc tests.





\subsection{Procedure}

\mytag{R2C9}\rev{For each optimizer}, including MMO paired with every choice/form of the auxiliary objectives, we use all \mytag{R3C1}\rev{212} cases from the SE optimization problems for the RQs. To ensure statistical validation, each experiment is repeated 10 times. All experiments are conducted in parallel on two high-performance servers, each with Ubuntu 20.04.6 LTS, two Intel(R) Xeon(R) Platinum 50-core CPUs at 2.30GHz, two Nvidia A100 80GB GPUs, and 500GB memory.




\section{Results and Analysis}
\label{sec:results}


\subsection{RQ1: Necessity of Best Auxiliary Objective Choice in MMO}
\label{sec:rq1}

\subsubsection{Method}

To confirm the necessity of choosing the best auxiliary objective for MMO, for each case, we compare the results of MMO paired with the best auxiliary objective (MMO\textsubscript{b}), selected from the seven choices considered in this work, against the original MMO\textsubscript{o} and {PMO}.


\definecolor{BlueCell}{RGB}{210,230,255}

\begin{table*}[t!]
\centering
\scriptsize

\caption{Comparing  MMO\textsubscript{b} with \texttt{SOTA} and \texttt{RS} across all cases and over 10 runs. The \rev{format} is the same as Table~\ref{tb:rq1_result}. Per case results can be found at: {\color{blue!60!black}\url{https://github.com/ideas-labo/mmo-sbse/blob/main/RQS/RQ2/RQ2.pdf}}.}
\label{tb:rq2_result}

\begin{adjustbox}{width=\linewidth,center}
\begin{tabular}{
l
>{\centering\arraybackslash}p{0.6cm} >{\centering\arraybackslash}p{0.6cm} >{\centering\arraybackslash}p{0.6cm}
>{\centering\arraybackslash}p{1.6cm} >{\centering\arraybackslash}p{1.6cm} >{\centering\arraybackslash}p{1.6cm}
>{\centering\arraybackslash}p{0.6cm} >{\centering\arraybackslash}p{0.6cm} >{\centering\arraybackslash}p{0.6cm}
>{\centering\arraybackslash}p{0.6cm} >{\centering\arraybackslash}p{0.6cm} >{\centering\arraybackslash}p{0.6cm}
}
\toprule
\multirow{2}{*}{\textbf{SE Optimization Problem}} &
\multicolumn{3}{c}{\textbf{Overall Rank}} &
\multicolumn{3}{c}{\textbf{Mean ± Std}} &
\multicolumn{3}{c}{\textbf{Ranked Best/Total}} &
\multicolumn{3}{c}{\textbf{Ranked Worst/Total}} \\

\cmidrule(lr){2-4} \cmidrule(lr){5-7} \cmidrule(lr){8-10} \cmidrule(lr){11-13}
& \textbf{MMO\textsubscript{b}} & \textbf{Baseline} & \textbf{RS} 
& \textbf{MMO\textsubscript{b}} & \textbf{Baseline} & \textbf{RS} 
& \textbf{MMO\textsubscript{b}} & \textbf{Baseline} & \textbf{RS} 
& \textbf{MMO\textsubscript{b}} & \textbf{Baseline} & \textbf{RS} \\
\midrule

Next Release Problem (NRP) &
\cellcolor{BlueCell}1.000 & 2.000 & 3.000 &
\cellcolor{BlueCell}$0.002 \pm 0.002$ & $0.412 \pm 0.036$ & $0.972 \pm 0.021$ &
\cellcolor{BlueCell}17/17 & 0/17 & 0/17 &
\cellcolor{BlueCell}0/17 & \cellcolor{BlueCell}0/17 & 17/17 \\

Software Configuration Tuning (SCT) &
\cellcolor{BlueCell}1.190 & 1.960 & 2.040 &
\cellcolor{BlueCell}$0.148 \pm 0.101$ & $0.280 \pm 0.173$ & $0.424 \pm 0.229$ &
\cellcolor{BlueCell}21/26 & 5/26 & 6/26 &
\cellcolor{BlueCell}1/26 & 17/26 & 17/26 \\

Software Effort Estimation (SEE) &
1.600 & \cellcolor{BlueCell}1.400 & 2.600 &
\cellcolor{BlueCell}$0.081 \pm 0.055$ & $0.110 \pm 0.081$ & $0.674 \pm 0.162$ &
6/15 & \cellcolor{BlueCell}9/15 & 1/15 &
4/15 & \cellcolor{BlueCell}1/15 & 14/15 \\

Software Defect Prediction (SDP) &
\cellcolor{BlueCell}1.020 & 1.550 & 2.120 &
\cellcolor{BlueCell}$0.214 \pm 0.100$ & $0.552 \pm 0.236$ & $0.540 \pm 0.231$ &
\cellcolor{BlueCell}39/40 & 24/40 & 3/40 &
\cellcolor{BlueCell}1/40 & 8/40 & 31/40 \\

Software Product Lines Testing (SPLT) &
\cellcolor{BlueCell}1.000 & 2.240 & 2.670 &
\cellcolor{BlueCell}$0.066 \pm 0.044$ & $0.560 \pm 0.126$ & $0.744 \pm 0.105$ &
\cellcolor{BlueCell}21/21 & 0/21 & 1/21 &
\cellcolor{BlueCell}0/21 & 6/21 & 15/21 \\

Software Project Scheduling (SPS) &
\cellcolor{BlueCell}1.000 & 2.000 & 3.000 &
\cellcolor{BlueCell}$0.004 \pm 0.003$ & $0.111 \pm 0.032$ & $0.973 \pm 0.020$ &
\cellcolor{BlueCell}18/18 & 0/18 & 0/18 &
\cellcolor{BlueCell}0/18 & 0/18 & 18/18 \\

Third Party Library Migration (TPLM) &
\cellcolor{BlueCell}1.310 & 1.570 & 2.110 &
\cellcolor{BlueCell}$0.139 \pm 0.146$ & $0.240 \pm 0.202$ & $0.565 \pm 0.100$ &
\cellcolor{BlueCell}24/35 & 15/35 & 11/35 &
11/35 & \cellcolor{BlueCell}5/35 & 24/35 \\

Workflow Scheduling (WS) &
\cellcolor{BlueCell}1.250 & 2.150 & 2.550 &
\cellcolor{BlueCell}$0.175 \pm 0.108$ & $0.403 \pm 0.233$ & $0.560 \pm 0.120$ &
\cellcolor{BlueCell}15/20 & 5/20 & 0/20 &
\cellcolor{BlueCell}1/20 & 8/20 & 12/20 \\

Web Service Composition (WSC) &
1.900 & \cellcolor{BlueCell}1.200 & 2.450 &
$0.530 \pm 0.239$ & \cellcolor{BlueCell}$0.403 \pm 0.227$ & $0.690 \pm 0.144$ &
6/20 & \cellcolor{BlueCell}17/20 & 0/20 &
7/20 & \cellcolor{BlueCell}1/20 & 15/20 \\

\midrule[0.8pt]
\mytag{R3C1}\rev{\textbf{Average}} &
\cellcolor{BlueCell}\rev{\textbf{1.252}} & \rev{1.786} & \rev{2.504} &
\cellcolor{BlueCell}\rev{$\mathbf{0.151 \pm 0.089}$} & \rev{$0.341 \pm 0.150$} & \rev{$0.682 \pm 0.126$} &
\cellcolor{BlueCell}\rev{\textbf{167/212}} & \rev{75/212} & \rev{22/212} &
\cellcolor{BlueCell}\rev{\textbf{25/212}} & \rev{46/212} & \rev{163/212} \\

\bottomrule
\end{tabular}
\end{adjustbox}
\end{table*}

\definecolor{BlueCell}{RGB}{210,230,255}

\begin{table*}[t!]
\centering
\scriptsize

\caption{Comparing  MMO\textsubscript{b} with other general-purpose optimizers in SE across all cases and over 10 runs. The \rev{format} is the same as Table~\ref{tb:rq1_result}. Per case results can be found at: {\color{blue!60!black}\url{https://github.com/ideas-labo/mmo-sbse/blob/main/RQS/RQ3/RQ3.pdf}}.}

\label{tb:rq3_result}

\begin{adjustbox}{width=\linewidth,center}
\begin{tabular}{
l
>{\centering\arraybackslash}p{0.6cm} >{\centering\arraybackslash}p{0.6cm} >{\centering\arraybackslash}p{0.6cm} >{\centering\arraybackslash}p{0.6cm}
>{\centering\arraybackslash}p{1.6cm} >{\centering\arraybackslash}p{1.6cm} >{\centering\arraybackslash}p{1.6cm} >{\centering\arraybackslash}p{1.6cm}
>{\centering\arraybackslash}p{0.6cm} >{\centering\arraybackslash}p{0.6cm} >{\centering\arraybackslash}p{0.6cm} >{\centering\arraybackslash}p{0.6cm}
>{\centering\arraybackslash}p{0.6cm} >{\centering\arraybackslash}p{0.6cm} >{\centering\arraybackslash}p{0.6cm} >{\centering\arraybackslash}p{0.6cm}
}
\toprule

\multirow{2}{*}{\textbf{SE Optimization Problem}} & 
\multicolumn{4}{c}{\textbf{Overall Rank}} &
\multicolumn{4}{c}{\textbf{Mean ± Std}} &
\multicolumn{4}{c}{\textbf{Ranked Best/Total}} &
\multicolumn{4}{c}{\textbf{Ranked Worst/Total}} \\

\cmidrule(lr){2-5} \cmidrule(lr){6-9} \cmidrule(lr){10-13} \cmidrule(lr){14-17}
& MMO\textsubscript{b} & SWAY & LINE & LITE
& MMO\textsubscript{b} & SWAY & LINE & LITE
& MMO\textsubscript{b} & SWAY & LINE & LITE
& MMO\textsubscript{b} & SWAY & LINE & LITE \\
\midrule

Next Release Problem (NRP) &
\cellcolor{BlueCell}1.000 & 2.290 & 3.650 & 2.180 &
\cellcolor{BlueCell}$0.002 \pm 0.002$ & $0.922 \pm 0.022$ & $0.961 \pm 0.028$ & $0.921 \pm 0.024$ &
\cellcolor{BlueCell}17/17 & 0/17 & 0/17 & 5/17 &
\cellcolor{BlueCell}0/17 & 1/17 & 16/17 & \cellcolor{BlueCell}0/17 \\

Software Configuration Tuning (SCT) &
\cellcolor{BlueCell}1.310 & 2.080 & 2.270 & 2.540 &
\cellcolor{BlueCell}$0.108 \pm 0.098$ & $0.325 \pm 0.197$ & $0.362 \pm 0.229$ & $0.444 \pm 0.229$ &
\cellcolor{BlueCell}20/26 & 6/26 & 7/26 & 5/26 &
\cellcolor{BlueCell}4/26 & 7/26 & 10/26 & 13/26 \\

Software Effort Estimation (SEE) &
\cellcolor{BlueCell}1.000 & 2.870 & 2.270 & 1.730 &
\cellcolor{BlueCell}$0.068 \pm 0.053$ & $0.673 \pm 0.180$ & $0.622 \pm 0.142$ & $0.625 \pm 0.143$ &
\cellcolor{BlueCell}15/15 & 0/15 & 1/15 & 7/15 &
\cellcolor{BlueCell}0/15 & 11/15 & 4/15 & 1/15 \\

Software Defect Prediction (SDP) &
\cellcolor{BlueCell}1.020 & 2.150 & 1.930 & 1.650 &
\cellcolor{BlueCell}$0.205 \pm 0.093$ & $0.519 \pm 0.217$ & $0.526 \pm 0.232$ & $0.521 \pm 0.226$ &
\cellcolor{BlueCell}39/40 & 3/40 & 16/40 & 26/40 &
\cellcolor{BlueCell}1/40 & 22/40 & 14/40 & 10/40 \\

Software Product Lines Testing (SPLT) &
\cellcolor{BlueCell}1.050 & 3.430 & 1.950 & 3.100 &
\cellcolor{BlueCell}$0.045 \pm 0.030$ & $0.746 \pm 0.120$ & $0.431 \pm 0.061$ & $0.725 \pm 0.112$ &
\cellcolor{BlueCell}20/21 & 0/21 & 1/21 & 0/21 &
\cellcolor{BlueCell}0/21 & 14/21 & \cellcolor{BlueCell}0/21 & 7/21 \\

Software Project Scheduling (SPS) &
\cellcolor{BlueCell}1.000 & 2.220 & 3.610 & 2.000 &
\cellcolor{BlueCell}$0.004 \pm 0.003$ & $0.912 \pm 0.019$ & $0.969 \pm 0.022$ & $0.914 \pm 0.018$ &
\cellcolor{BlueCell}18/18 & 0/18 & 0/18 & 7/18 &
\cellcolor{BlueCell}0/18 & \cellcolor{BlueCell}0/18 & 18/18 & \cellcolor{BlueCell}0/18 \\

Third Party Library Migration (TPLM) &
\cellcolor{BlueCell}1.090 & 2.630 & 2.140 & 1.660 &
\cellcolor{BlueCell}$0.155 \pm 0.133$ & $0.668 \pm 0.117$ & $0.614 \pm 0.108$ & $0.614 \pm 0.128$ &
\cellcolor{BlueCell}33/35 & 7/35 & 11/35 & 16/35 &
\cellcolor{BlueCell}1/35 & 21/35 & 6/35 & \cellcolor{BlueCell}1/35 \\

Workflow Scheduling (WS) &
\cellcolor{BlueCell}1.000 & 2.400 & 2.850 & 1.850 &
\cellcolor{BlueCell}$0.176 \pm 0.143$ & $0.637 \pm 0.137$ & $0.693 \pm 0.172$ & $0.627 \pm 0.162$ &
\cellcolor{BlueCell}20/20 & 0/20 & 3/20 & 9/20 &
\cellcolor{BlueCell}0/20 & 6/20 & 13/20 & 4/20 \\

Web Service Composition (WSC) &
\cellcolor{BlueCell}1.300 & 2.500 & 1.750 & 1.950 &
\cellcolor{BlueCell}$0.458 \pm 0.253$ & $0.651 \pm 0.167$ & $0.621 \pm 0.147$ & $0.638 \pm 0.173$ &
\cellcolor{BlueCell}17/20 & 2/20 & 7/20 & 8/20 &
\cellcolor{BlueCell}3/20 & 13/20 & 4/20 & 6/20 \\

\midrule[0.8pt]
\textbf{Average} &
\cellcolor{BlueCell}\textbf{1.086} & 2.508 & 2.491 & 2.073 &
\cellcolor{BlueCell}$\mathbf{0.136 \pm 0.090}$ & $0.673 \pm 0.131$ & $0.644 \pm 0.127$ & $0.670 \pm 0.135$ &
\cellcolor{BlueCell}\textbf{199/212} & 18/212 & 46/212 & 83/212 &
\cellcolor{BlueCell}\textbf{9/212} & 95/212 & 85/212 & 42/212 \\

\bottomrule
\end{tabular}
\end{adjustbox}
\end{table*}

\subsubsection{Results}
As shown in Table~\ref{tb:rq1_result}, MMO\textsubscript{b} achieves the best overall performance across all \mytag{R3C1}\rev{nine} SE optimization problems. In terms of Scott--Knott ESD ranks, MMO\textsubscript{b} attains an overall mean rank of \mytag{R3C1}\rev{1.079}, outperforming MMO\textsubscript{o} (\mytag{R3C1}\rev{1.573}) and {PMO} (\mytag{R3C1}\rev{2.126}), indicating its statistically consistent advantage. This trend is also reflected in the normalized target objective values: MMO\textsubscript{b} achieves a mean normalized performance of \mytag{R3C1}\rev{$0.198 \pm 0.128$}, significantly better than MMO\textsubscript{o} (\mytag{R3C1}\rev{$0.266 \pm 0.142$}) and {PMO} (\mytag{R3C1}\rev{$0.570 \pm 0.193$}). At the case level, MMO\textsubscript{b} achieves the best rank in \mytag{R3C1}\rev{198 out of 212 cases (93\%)} and the worst rank in only \mytag{R3C1}\rev{10 cases (5\%)}. In comparison, MMO\textsubscript{o} attains the best rank in \mytag{R3C1}\rev{104 cases (49\%)} and the worst rank in \mytag{R3C1}\rev{49 cases (23\%)}, while {PMO} achieves the best rank in \mytag{R3C1}\rev{34 cases (16\%)} and the worst rank in \mytag{R3C1}\rev{160 cases (75\%)}.

These results indicate that MMO\textsubscript{b} provides \mytag{R2C1}\rev{a} more effective way of overcoming local optima by systematically leveraging the most suitable auxiliary objective for each SE optimization problem/case, demonstrating its suitability to serve as a general-purpose baseline therein.

Looking at individual SE optimization problems, this trend remains consistent in most cases, where MMO\textsubscript{b} is still \mytag{R2C1}\rev{superior to} the others, including SE problems that are known to be expensive and complex with rugged \mytag{R2C1}\rev{landscapes} (e.g., SCT) or those that involve an extremely intractable space (e.g., NRP).

\begin{tcbitemize}[%
    raster columns=1, 
    raster rows=1
    ]
  \tcbitem[myhbox={}{Finding \thefindingcount}]
  \textit{Compared with MMO\textsubscript{o} and {PMO}, MMO\textsubscript{b} attains the top average rank for all SE optimization problems (the best rank in \mytag{R3C1}\rev{93\%} and the worst merely in 5\% of the cases). This \mytag{R2C1}\rev{confirms} the necessity of choosing the best auxiliary objective for MMO on each case.}
\end{tcbitemize}
\addtocounter{findingcount}{1}

\subsection{RQ2: MMO\textsubscript{b} against State-of-the-art Optimizers}
\label{sec:rq2}

\subsubsection{Method}

Since we have empirically confirmed the importance of choosing the best auxiliary objective, next, we compare MMO\textsubscript{b} against the state-of-the-art optimizers in Table~\ref{tb:sbse_single_baselines} in order to assess the effectiveness of MMO\textsubscript{b} for individual SE optimization problems, denoted as \texttt{SOTA}, together with a Random Search (\texttt{RS}) baseline.




\subsubsection{Results}

As shown in Table~\ref{tb:rq2_result}, MMO\textsubscript{b} achieves the best overall performance across all SE optimization problems---an average Scott--Knott ESD rank of \mytag{R3C1}\rev{1.252}, which is significantly better than the state-of-the-art optimizer (\mytag{R3C1}\rev{1.786}) and random search (\mytag{R3C1}\rev{2.504}), suggesting that systematically selecting an appropriate auxiliary objective for each problem consistently guides MMO toward stable and strong optimization results. The normalized objective values reflect a similar trend: MMO\textsubscript{b} achieves an average performance of \mytag{R3C1}\rev{$0.151 \pm 0.089$}, outperforming the state-of-the-art optimizer (\mytag{R3C1}\rev{$0.341 \pm 0.150$}) and random search (\mytag{R3C1}\rev{$0.682 \pm 0.126$}). At the case level, MMO\textsubscript{b} attains the top rank in \mytag{R3C1}\rev{167 out of 212 cases (79\%)} and ranks last in only \mytag{R3C1}\rev{25 cases (12\%)}, which is substantially better than the others. These results indicate that systematically selecting the best auxiliary objective in MMO for each SE problem can provide a strong baseline.

A closer look at individual SE optimization problem reveals that MMO\textsubscript{b} exhibits pronounced advantages in most problems, particularly those with larger search spaces or more complex landscapes where local optima can be particularly problematic, such as SCT, NRP, and SPS. However, in two SE problems (i.e., WSC and SEE), MMO\textsubscript{b} performs slightly worse than the state-of-the-art optimizer. This is because (1) these SE problems have smoother objective landscapes, hence the benefit auxiliary objective is blurred; (2) the \texttt{SOTA} for cases of these problems is already close to the global optimum.

\begin{tcbitemize}[%
    raster columns=1, 
    raster rows=1
    ]
  \tcbitem[myhbox={}{Finding \thefindingcount}]
  \textit{MMO\textsubscript{b} achieves the best overall average rank and leads in \mytag{R3C1}\rev{79\%} of the cases, demonstrating clear and consistent superiority over different \texttt{SOTA} and \texttt{RS}, when the right auxiliary objective is chosen.}
\end{tcbitemize}
\addtocounter{findingcount}{1}

\subsection{RQ3: The Generality of MMO}

\subsubsection{Method}

Here, we seek to assess whether MMO, when paired with the best auxiliary objective, can have competitive performance compared with other general-purpose optimizers for SE optimization, i.e., \texttt{SWAY}, \texttt{LINE}, and \texttt{LITE}.


\subsubsection{Results}

As shown in Table~\ref{tb:rq3_result}, among the general-purpose optimizers, \texttt{SWAY} generally has the worst average ranks, indicating limited competitiveness. \texttt{LINE} achieves moderate performance, occasionally obtaining favorable results in certain problems, while \texttt{LITE} shows some variability, achieving slightly better average ranks than the other two in some SE tasks but remain limited in robustness. This trend can be attributed to the fact that, unlike the chosen auxiliary objective in MMO for a case, these general-purpose optimizers are not tailored to specific characteristics of the SE problems, limiting their ability to guide the search effectively across diverse SE optimization tasks.

In contrast, MMO\textsubscript{b} consistently outperforms the other optimizers in both average rank and normalized objective values. Regarding the number of best and worst cases, MMO\textsubscript{b} achieves the top rank in \mytag{R3C1}\rev{199 out of 212 cases ($94\%$)} and the worst rank in only \mytag{R3C1}\rev{9 cases ($4\%$)}. In comparison, \texttt{SWAY}, \texttt{LINE}, and \texttt{LITE} achieve the best rank in \mytag{R3C1}\rev{18, 46, and 83 cases}, and the worst rank in \mytag{R3C1}\rev{95, 85, and 42 cases}, respectively. These results indicate that the problem/case-specific auxiliary objectives adopted by MMO are effective in overcoming local optima and guiding the optimization, enabling it to consistently find high-quality solutions across nearly all cases.

When examining individual SE optimization problems, MMO\textsubscript{b} demonstrates particularly strong advantages in tasks with larger search spaces or more complex landscapes, such as the NRP---this is precisely the strength of MMO, especially under the most suitable auxiliary objective.

\begin{tcbitemize}[raster columns=1, raster rows=1]
  \tcbitem[myhbox={}{Finding \thefindingcount}]
  \textit{With the right auxiliary objective, MMO\textsubscript{b} is generally ranked the best in \mytag{R3C1}\rev{94\% of the 212 cases} with much improved results, consistently outperforming other general-purpose optimizers at all levels.}
\end{tcbitemize}
\addtocounter{findingcount}{1}

\begin{figure}[t!]
\centering
\begin{tikzpicture}
\begin{axis}[
    ybar stacked,
    width=17cm,
    height=12cm,
    ymin=0, ymax=100.01,
    bar width=1cm,
    grid=major,
    grid style={dashed,gray!30},
    ylabel={Percentage of proportion (\%)},
    ylabel style={font=\LARGE},
    xlabel={SE optimization problems},
    xlabel style={font=\LARGE},
    symbolic x coords={SDP, SPS, SPLT, TPLM, SEE, NRP, WSC, WS, SCT},
    xtick=data,
    xticklabel style={rotate=45, anchor=east, font=\Large, yshift=-2pt},
    enlarge x limits=0.06,
    nodes near coords,
    every node near coord/.style={
        font=\normalsize\bfseries,
        anchor=center,
        rotate=0,
        fill=white,
        fill opacity=0.9,
        text opacity=1,
        rounded corners=0.8pt,
        inner xsep=1.2pt,
        inner ysep=0.6pt
    },
    legend style={
        at={(0.5,1.0)},
        anchor=south,
        legend columns=7,
        draw=none,
        font=\Large,
        cells={anchor=center},
        align=center,
        yshift=8pt,
        /tikz/every even column/.append style={column sep=0.3cm}
    },
    tick label style={font=\Large},
    title style={font=\LARGE},
    x tick label style={yshift=-5pt}
]

\addplot[
    ybar,
    forget plot,
    fill=color1,
    draw=black,
    postaction={pattern=horizontal lines, pattern color=black},
    /pgfplots/legend image code/.code={
        \path[fill=color1] (0cm,-0.1cm) rectangle (0.4cm,0.1cm);
        \draw[black] (0cm,0cm) -- (0.4cm,0cm);
    }
] coordinates {
(SDP,0) (SPS,27.78) (SPLT,38.1) (TPLM,17.14) (SEE,33.33) (NRP,94.12) (WSC,25.0) (WS,15.0) (SCT,11.54)
};

\addplot[ybar, forget plot, fill=color2, draw=black, postaction={pattern=vertical lines, pattern color=black}] coordinates {
(SDP,0) (SPS,0) (SPLT,0) (TPLM,0) (SEE,0) (NRP,0) (WSC,15.0) (WS,5.0) (SCT,19.23)
};

\addplot[ybar, forget plot, fill=color3, draw=black, postaction={pattern=north east lines, pattern color=black}] coordinates {
(SDP,20.0) (SPS,33.33) (SPLT,0) (TPLM,25.71) (SEE,13.33) (NRP,0) (WSC,10.0) (WS,5.0) (SCT,3.85)
};

\addplot[ybar, forget plot, fill=color4, draw=black, postaction={pattern=north west lines, pattern color=black}] coordinates {
(SDP,5.0) (SPS,5.56) (SPLT,4.76) (TPLM,11.43) (SEE,20.0) (NRP,0) (WSC,0) (WS,0) (SCT,15.38)
};

\addplot[ybar, forget plot, fill=color5, draw=black, postaction={pattern=grid, pattern color=black}] coordinates {
(SDP,5.0) (SPS,0) (SPLT,42.86) (TPLM,8.57) (SEE,6.67) (NRP,0) (WSC,15.0) (WS,20.0) (SCT,19.23)
};

\addplot[ybar, forget plot, fill=color6, draw=black, postaction={pattern=crosshatch, pattern color=black}] coordinates {
(SDP,62.5) (SPS,27.78) (SPLT,9.52) (TPLM,28.57) (SEE,26.67) (NRP,5.88) (WSC,25.0) (WS,35.0) (SCT,15.38)
};

\addplot[ybar, forget plot, fill=color7, draw=black, postaction={pattern=dots, pattern color=black}] coordinates {
(SDP,7.5) (SPS,5.56) (SPLT,4.76) (TPLM,8.57) (SEE,0) (NRP,0) (WSC,10.0) (WS,20.0) (SCT,15.38)
};

\addlegendimage{/pgfplots/legend image code/.code={
    \path[fill=color1] (0cm,-0.1cm) rectangle (0.6cm,0.1cm);
    \draw[black, line width=0.8pt] (0cm,0cm) -- (0.6cm,0cm);
}}
\addlegendentry{Original}
\addlegendimage{area legend, fill=color2, draw=black, postaction={pattern=vertical lines, pattern color=black}}
\addlegendentry{Age}
\addlegendimage{area legend, fill=color3, draw=black, postaction={pattern=north east lines, pattern color=black}}
\addlegendentry{Gaussian Noise}
\addlegendimage{area legend, fill=color4, draw=black, postaction={pattern=north west lines, pattern color=black}}
\addlegendentry{Grid Diversity}
\addlegendimage{area legend, fill=color5, draw=black, postaction={pattern=grid, pattern color=black}}
\addlegendentry{Novelty}
\addlegendimage{area legend, fill=color6, draw=black, postaction={pattern=crosshatch, pattern color=black}}
\addlegendentry{Penalty}
\addlegendimage{area legend, fill=color7, draw=black, postaction={pattern=dots, pattern color=black}}
\addlegendentry{Reciprocal}

\end{axis}
\end{tikzpicture}
\caption{The proportion of the best auxiliary objective choice for MMO across all cases under each SE optimization problem.}

\label{fig:mode_percentage}
\end{figure}
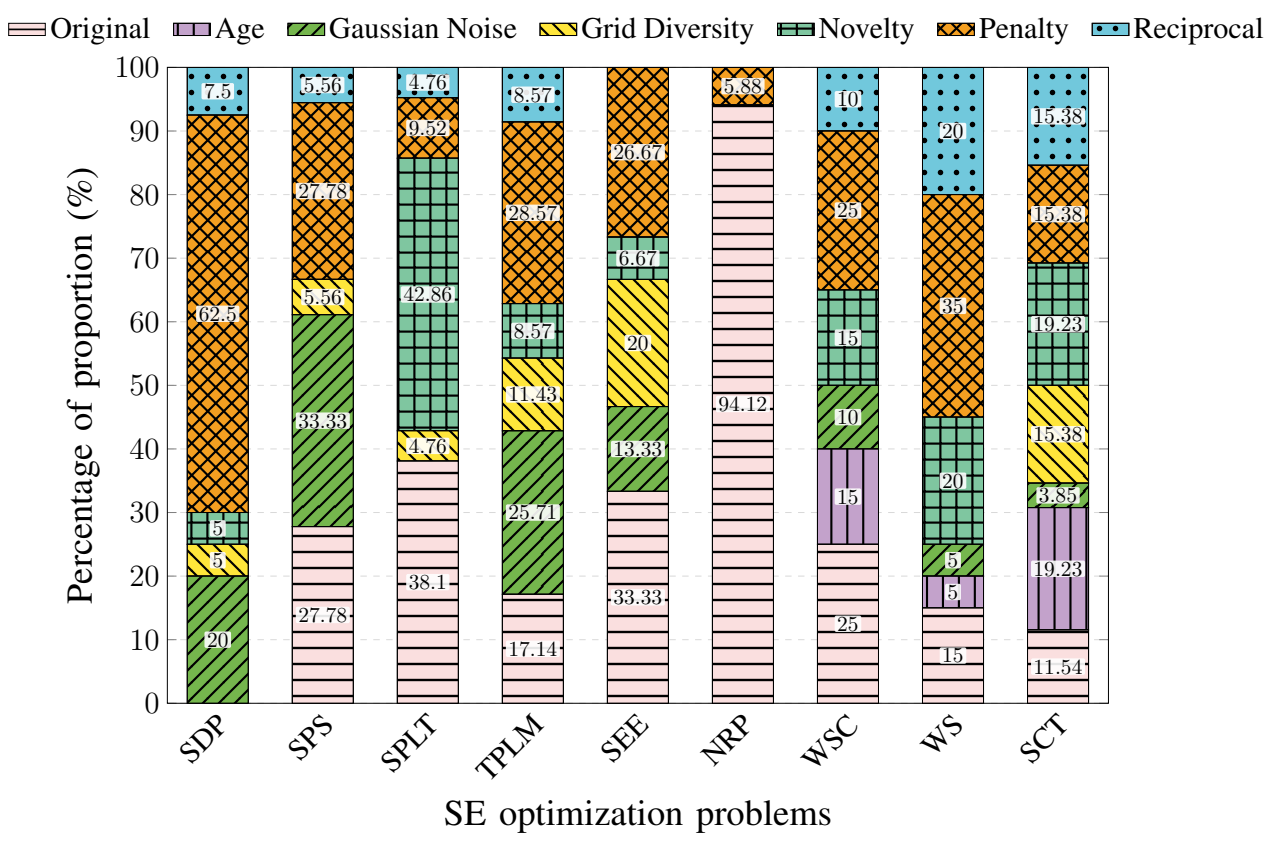

\subsection{Insight from the Findings}
\label{sec:discussion}

The findings from RQ1--RQ3 lead to a clear insight:
\begin{tcolorbox}[enhanced,boxrule=0.2mm, title=\textbf{\textit{Key Insight}},
colframe=black,colback=white,colbacktitle=white,
coltitle=black,attach boxed title to top center=
{yshift=-0.25mm-\tcboxedtitleheight/2,yshifttext=2mm-\tcboxedtitleheight/2},
boxed title style={boxrule=-0.2mm,
frame code={ \path[tcb fill frame] ([xshift=-3mm]frame.west)
-- (frame.north west) -- (frame.north east) -- ([xshift=3mm]frame.east)
-- (frame.south east) -- (frame.south west) -- cycle; },
interior code={ \path[tcb fill interior] ([xshift=-2.1mm]interior.west)
-- (interior.north west) -- (interior.north east)
-- ([xshift=2.1mm]interior.east) -- (interior.south east) -- (interior.south west)
-- cycle;} }]
MMO, even when paired with a fairly standard optimizer for SBSE (i.e., NSGA-II), can serve as a strong baseline for SE optimization, provided that the most suitable auxiliary objective can be chosen.
\end{tcolorbox}

The above is promising, especially considering the simplicity that MMO offers. However, the strong performance of MMO relies on one condition: the best auxiliary objective has been chosen---if that ``best'' is consistent across the SE optimization problems/cases, then we can reliably recommend one, or otherwise there is an extra level of complexity.

To verify this, Figure~\ref{fig:mode_percentage} shows which of the auxiliary objective choices are the best across all the cases of SE optimization problems. \mytag{R3C4}\rev{Unfortunately, we see that the ``best'' can be rather diverse depending on the case: it is difficult to state which is generally the best even for all cases within most of the same SE optimization problem, let alone identifying a generally best one across all SE problems. In particular, there is no single auxiliary objective choice constantly having a large proportion of ``best'' over all problems. Rather, the results are diverse; for example, \textit{Original} is the best for $94.12\%$ of cases for NRP, but has $0\%$ wins for SDP. Even when considering different cases within an individual SE problem, NRP is the only problem where we can confidently say there exists a predominantly best auxiliary objective choice ($94.12\%$ of cases); for the others, the most common auxiliary objective choice ranges between $19.23\%$ and $62.5\%$.}

\mytag{R3C4}\rev{Thus, the above leaves us with a question on specializing MMO for a given SE optimization problem/case: how to choose the best auxiliary objective for MMO without expensive trial-and-error for specializing it as a strong baseline?}

\begin{table*}[t!]
\centering
\footnotesize
\caption{Categories and brief descriptions of prediction features.}
\label{tab:landscape_features}
\begin{adjustbox}{width=0.8\linewidth,center}
\begin{tabular}{p{5.2cm} p{10.6cm}}
\toprule
\textbf{Feature} & \textbf{Description} \\ 
\midrule
\rowcolor{lightgray}\multicolumn{2}{c}{\textbf{Single-objective Landscape Features}} \\
\midrule

Skewness (Ske) &
Measures the asymmetry of the fitness value distribution over the solution set. \\

Kurtosis (Kur) &
Measures the peakedness and concentration of the fitness value distribution. \\

Maximum Information Entropy (MIE) &
Quantifies the information complexity and heterogeneity of solutions in the decision space. \\

Nearest Better Clustering (NBC) &
Characterizes the spatial clustering tendency of better solutions relative to their neighbors. \\

\midrule
\rowcolor{lightgray}\multicolumn{2}{c}{\textbf{Dominance-based Features}} \\
\midrule

Pareto Dominated Solution Ratio (PDSR) &
Proportion of solutions that are Pareto-dominated by at least one other solution. \\

Level Count (LC) &
Number of non-dominated fronts obtained through Pareto-based sorting of the solution set. \\

Local Efficient Points Count (LEPC) &
Number of locally Pareto-efficient solutions in the decision space. \\

Local-to-Global Ratio (LGR) &
Ratio of locally Pareto-efficient solutions that are also globally Pareto-optimal. \\

Connected Components (CC) &
Connectivity structure among locally efficient solutions in the decision space. \\

Connected Components (Pareto) (CC-P) &
Connectivity structure among globally Pareto-optimal solutions. \\

Maximum Dominance Count (MDC) &
Maximum number of solutions dominated by a single solution. \\

Median Dominance Count (MedDC) &
Median number of solutions dominated across the solution set. \\

Dominance Change Frequency (DCF) &
Frequency of dominance relationship changes within local neighborhoods. \\

Average Dominance Pressure (ADP) &
Average local dominance pressure exerted on solutions by their neighbors. \\

Hypervolume (HV) &
Volume of the objective space weakly dominated by the Pareto solution set. \\

Pareto Grid Ratio (PGR) &
Ratio of occupied grid cells by Pareto-optimal solutions in the discretized objective space. \\

\midrule
\rowcolor{lightgray}\multicolumn{2}{c}{\textbf{Objective-space Distribution Features}} \\
\midrule

Spearman Correlation Coefficient (SCC) &
Measures rank-based correlations between pairs of objectives. \\

Pearson Correlation Coefficient (PCC) &
Measures linear correlations between pairs of objectives. \\

Kendall's Tau (KTC) &
Measures ordinal consistency between objective rankings. \\

Slope of Linear Regression (SLR) &
Slope of the linear regression model fitted between two objectives. \\

Objective Function Value Variance (OFV) &
Quantifies the dispersion of objective values across the solution set. \\

Objective Function Value Entropy (OFE) &
Measures the entropy of objective value distributions in the objective space. \\

Average Distance (AD) &
Average pairwise distance among Pareto-optimal solutions, reflecting diversity. \\

Grid Density Variance (GDV) &
Variance of Pareto solution density across objective-space grids. \\

Adjacent Grid Distance Mean (AGDM) &
Average distance between adjacent grid cells, characterizing local smoothness. \\

\bottomrule
\end{tabular}
\end{adjustbox}
\end{table*}

\begin{table*}[t!]
\centering
\footnotesize
\caption{Sampling methods used for each SE optimization problem for collecting the training data of \approach.}


\label{tb:sbse_sampling}

\begin{adjustbox}{width=0.8\linewidth,center}
\begin{tabular}{
l
l
}
\toprule
\textbf{SE Optimization Problem} & \textbf{Sampling Methods} \\
\midrule

\rowcolor{BlueCell}Next Release Problem (NRP) & Monte Carlo, Sobol Sequence, Latin Hypercube, Orthogonal, Halton Sequence, Covering Array \\

Software Configuration Tuning (SCT) & Monte Carlo, Sobol Sequence, Latin Hypercube, Orthogonal, Stratified, Covering Array \\

\rowcolor{BlueCell}Software Effort Estimation (SEE) & Monte Carlo, Sobol Sequence, Latin Hypercube, Stratified, Halton Sequence, Random Walk \\

Software Defect Prediction (SDP) & Monte Carlo, Sobol Sequence, Latin Hypercube, Orthogonal, Stratified, Covering Array \\

\rowcolor{BlueCell}Software Product Lines Testing (SPLT) & Monte Carlo, Sobol Sequence, Latin Hypercube, Halton Sequence, Stratified, Covering Array \\

Software Project Scheduling (SPS) & Monte Carlo, Sobol Sequence, Latin Hypercube, Stratified, Halton Sequence, Random Walk \\

\rowcolor{BlueCell}Third Party Library Migration (TPLM) & Monte Carlo, Sobol Sequence, Latin Hypercube, Orthogonal, Stratified, Covering Array \\

Workflow Scheduling (WS) & Monte Carlo, Sobol Sequence, Latin Hypercube, Orthogonal, Stratified, Covering Array \\

\rowcolor{BlueCell}Web Service Composition (WSC) & Monte Carlo, Sobol Sequence, Latin Hypercube, Orthogonal, Stratified, Covering Array \\

\bottomrule
\end{tabular}
\end{adjustbox}
\end{table*}

\section{Predicting the Choice/Form of Auxiliary Objective with \approach}
\label{sec:tool}

\subsection{\mytag{R2C2}\rev{Motivation}}

\mytag{R2C2}\rev{The insight from our empirical study confirms the importance of choosing a suitable auxiliary objective for MMO when it is applied to a case of an SE optimization problem. There are three straightforward ways to make this choice:}

\begin{itemize}
   \item \textbf{\rev{Choosing the best one via exhaustive trial-and-error:}} \rev{One has to run MMO under every candidate auxiliary-objective choice to obtain the ranks, as what we have done for constructing the ground truth in this work.}
    \item \textbf{\rev{Randomly choosing one:}} \rev{This incurs no overhead, but the choice is not supported by any evidence.}
    \item \textbf{\rev{Choosing the generally best one from our empirical study:}} \rev{This appears reasonable, but our results show that it is difficult to identify a generally dominating auxiliary objective across all SE problems and cases.}
\end{itemize}

\rev{Clearly, none of the above is ideal: trial-and-error is accurate but expensive; random choice can lead to poor results; and using a generally best choice may still fail for individual cases/SE problem. Therefore, what we need is an automated approach that can select a suitable auxiliary objective for MMO with reasonable accuracy and efficiency.}

To assist the choice of a suitable auxiliary objective for using MMO as the baseline for SE optimization, we propose \approach, a method that predicts the best auxiliary objective choice, given a new, unforeseen case of the SE problem.

\subsection{Problem Formulation}
\label{sec:approach}

We formulate choosing the auxiliary objective for MMO under a case of SE optimization as a learning-to-rank problem:
\begin{equation}
\min \sum_{k=1}^{K} \sum_{i=1}^{n_k} \mathcal{L}(\mathbf{F}_{k,i}, r_{k,i})
\label{eq:ltr}
\end{equation}
where $\mathbf{F}_{k,i}$ denotes some features related to the chosen auxiliary objective, MMO, and the problem characteristics, for the $i$th candidate auxiliary objective in the $k$th case; $r_{k,i}$ represents its observed rank according to MMO performance, and $\mathcal{L}$ is the ranking loss function to be minimized. Here, we consider all seven choices of auxiliary objective studied. Our goal is to minimize the loss such that, for a newly given case with some sampled data, the margin between the predicted rank of an auxiliary objective and the ground truth is minimal.

\subsection{Features for Distinguishing Auxiliary Objectives}

To address Equation~\ref{eq:ltr}, we design three feature types based on sampled solutions and target/auxiliary objective values:

\begin{itemize}
    \item \textbf{Objective-space Distribution Feature:} 
    Based on the theoretical analysis in Section~\ref{first_motivation}, the effectiveness of an auxiliary objective for MMO critically depends on whether it satisfies \textbf{Property~3}. Accordingly, we use several objective-space distribution measures to quantify the relationship between the target and the selected auxiliary objective, as well as the distribution of their induced Pareto-optimal solutions in the objective space, e.g., the Pearson correlation, Kendall's Tau, objectives' entropy.


    \item \textbf{Dominance-based Feature:} This refers to the dominance relationships between $g_1$ and $g_2$ in the MMO space after rotation. The extent of comparability, determined by the auxiliary objective, largely influences the performance of MMO. We use common metrics such as Pareto dominance solution ratio, level count, etc. 

    \item \textbf{Landscape Feature:} This type of feature describes the problem characteristics, measured by landscape metrics from the well-established fitness landscape analysis~\cite{liefooghe2019landscape, DBLP:conf/gecco/LiefoogheVLZM21, ye2026revealing,DBLP:conf/gecco/MersmannBTPWR11,pearson1895contributions}, which directly quantifies the difficulty and extent of local optima for the problem. Common ones are Kurtosis and nearest better clustering. 
\end{itemize}

Since the actual metrics for each type are vast, we conduct an information-theory based feature selection and identify 25 top relevant features for the SE optimization problems, as in Table~\ref{tab:landscape_features}. More details of the features are provided in Appendix C.

\subsection{Training Procedure}

\begin{figure}[t!]
\centering
\includegraphics[width=\columnwidth]{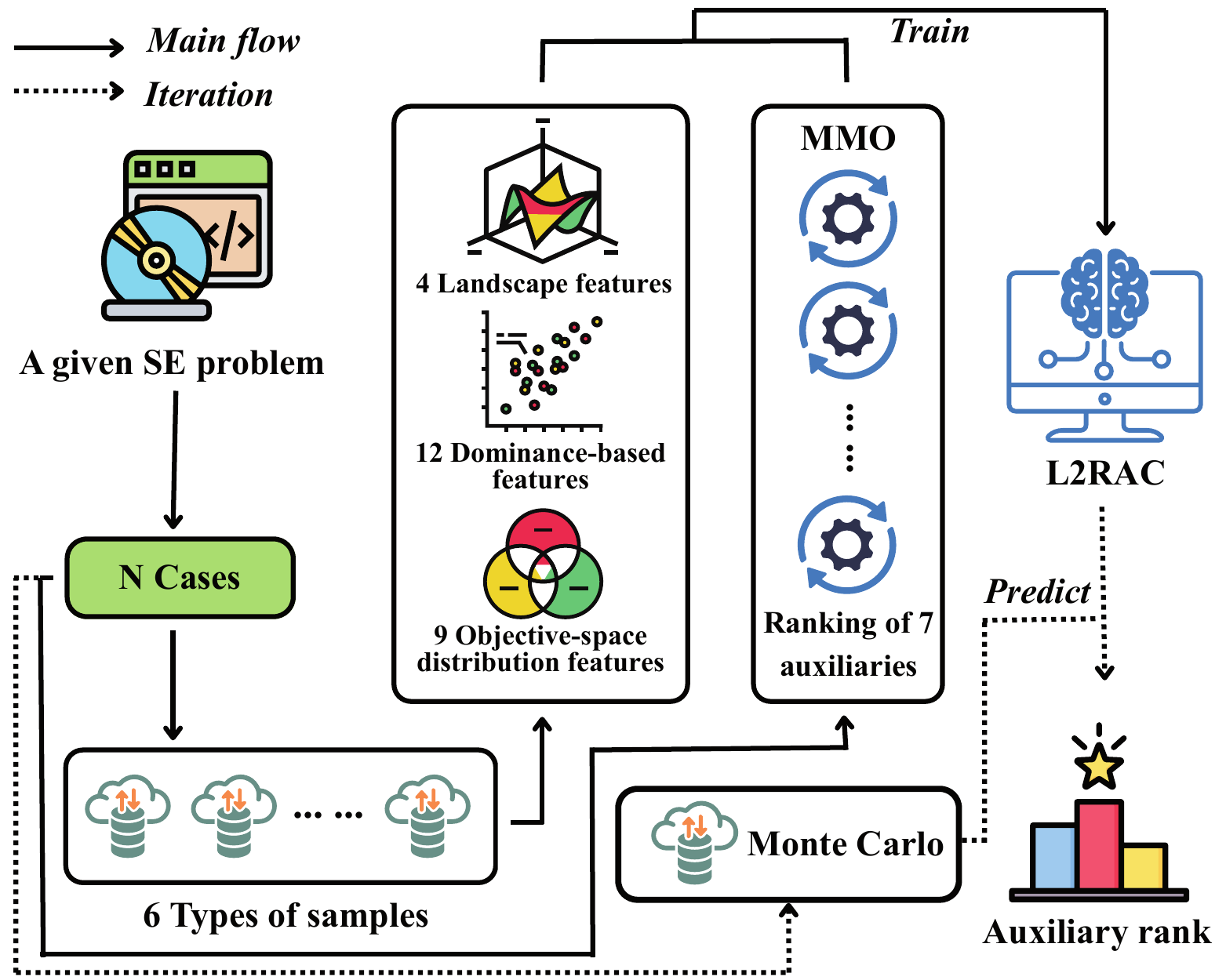}
\caption{Overview of \approach.}
\label{fig:workflow}
\end{figure}

As illustrated in Figure~\ref{fig:workflow}, \approach~relies on \texttt{LambdaRank} \cite{DBLP:conf/nips/KeMFWCMYL17} as the core to build a model in prediction. In particular, \approach~is a kind of problem specific pre-training model: it can be trained on multiple known cases of a SE optimization problem, and the trained model can then be applied to an unforeseen case of that SE problem. Therefore, given a SE problem, the training can follow the procedure outlined below:

\begin{enumerate}[label=(\roman*)]

    \item \textbf{Case Selection.} Select a case $\mathcal{C}$ from the SE problem.
    
    \item \textbf{Performance Collection.} For $\mathcal{C}$, run optimization using all the possible auxiliary objective choices with MMO over 10 runs\footnote{Budget can be set depending on the resource; in this work, we use the same budget as the empirical study.}. Those choices can then be ranked based on Scott-Knott ESD test, the median performance of the target objective, and then complexity---the same process we used when determining MMO\textsubscript{b}---as the ground truth.
    
    
    \item \textbf{Sampling.} Sample solutions using multiple sampling strategies defined for the corresponding SE problem (see Table~\ref{tb:sbse_sampling}) on $\mathcal{C}$, in which the sampling methods are chosen from the widely-used ones for the specific SE problem. Each sampling method is repeated under 10 different random seeds to ensure robustness and coverage, leading to 10 datasets. For the sample size, in this work, we cap as $1,000$ solutions per sampling method.
    
    \item \textbf{Feature Computation.} For every dataset, all sampled solutions can be used to compute the feature values in Table~\ref{tab:landscape_features}, yielding 10 feature vectors per sampling method. To mitigate random variation, the median values across the 10 repetitions are used as the feature values of $\mathcal{C}$ under a corresponding sampling strategy.

    \item Repeat from (i) until all cases are covered.



    \item \textbf{Model Training.} With all the feature values and the ground truth of auxiliary objective ranking for each of the cases from all the SE problems (removing redundant ones), we train the model via \texttt{LambdaRank} using all the data from the collected feature values and ranks.

\end{enumerate}

\mytag{R1C3}\rev{The sampling setting differs between model training and prediction. During model training, we use multiple sampling strategies across the studied SE problems/cases, because different samplers provide different coverage patterns and reduce dependence on a single sampling bias. During prediction on a new case, however, \approach{} does not require users to repeat all these sampling strategies for the simplicity of application. Once the ranking model has been trained, prediction can be instantiated with one sampling method depending on the preference; in this work, we use Monte Carlo sampling as the default choice. Other more advanced samplers, such as those used for training, can also be used as alternatives.}

Note that it is important to reorder the sampled solutions in a way to approximate the optimization convergence, as required by certain auxiliary objectives, i.e., \textit{Age}, \textit{Diversity}, \textit{Penalty}, \textit{Novelty}. For example, the \textit{Age} depends on the chronological order of the explored solution, hence we sort the solutions from the worst target objective value to the best.





\subsection{Prediction}

\mytag{R3C5}\rev{Once trained, \approach~can predict for a new, unforeseen case of the SE problem by taking some samples to compute the features. The outcome is the rank of all seven auxiliary objectives for MMO, where the top one is the estimated best.}

\subsection{Evaluation}

\subsubsection{Metric}

To assess \approach, we use Normalized Discounted Cumulative Gain (NDCG)---a common evaluation metric used in learning to rank. For a predicted ranking of length $k$, the Discounted Cumulative Gain (DCG) is
\begin{equation}
\mathrm{DCG}_k = \sum_{i=1}^{k}\frac{2^{rel_i}}{\log_2(i+1)}
\end{equation}
where $rel_i$ is the relevance of the item at position $i$ derived from the true ranking based on the target objective. The normalized DCG is then defined as:
\begin{equation}
\mathrm{NDCG}_k = \frac{\mathrm{DCG}_k}{\mathrm{IDCG}_k}
\end{equation}
where $\mathrm{IDCG}_k$ denotes the ideal DCG obtained from the ground-truth ranking for the top $k$ items. NDCG ranges from 0 to 1, with higher values are preferred.

We report on NDCG@1 and NDCG@3 for measuring the consistency with top-1 and top-3 ranks in the ground truth\footnote{Note that the ground truth in NDCG@1 is consistent with the MMO\textsubscript{b} used in our empirical study}, respectively, and NDCG that is computed over the full ranking. 

\subsubsection{Procedure}

For each SE optimization problem, we adopt the leave-one-out strategy in the evaluation: \approach~is trained on data collected from $n-1$ cases, leaving one case out for testing. We repeat the same until every case serves as the testing case once, and compute their corresponding NDCG values. For example, there would be 26 NDCG values in the comparisons for the SCT, which has 26 cases. Note that some samples are required for the new problem case at prediction (we use $1,000$ solutions collected by Monte Carlo sampling as the default). For all experiments, we repeat 10 runs.

\subsubsection{Results}

\begin{table}[t!]
\centering
\definecolor{BlueCell}{RGB}{210,230,255}
\caption{
Average NDCG performance of \approach~over random ranking. 
The cell value $\alpha^{\pm\beta\%}$ means \approach~achieves $\alpha$ on NDCG, 
which is $+\beta\%$ better (or $-\beta\%$ worse) than random guessing. 
Here \colorbox{BlueCell}{blue cells} indicate a positive and statistically significant improvement, 
while white cells indicate no significant improvement. 
$^\dagger$ and $^\star$ denote $p < 0.001$ and $0.001 \leq p < 0.05$, respectively. 
Detailed case-level results for each SE problem can be found at: 
{\color{blue!60!black}\url{https://github.com/ideas-labo/mmo-sbse/blob/main/RQS/predict/predict.pdf}}
}

\label{tab:folder_ndcg_redbetter}
\begin{adjustbox}{width=0.95\linewidth,center}
\begin{tabular}{l|lll}
\toprule
\textbf{SE Optimization Problem} & \textbf{NDCG@1} & \textbf{NDCG@3} & \textbf{NDCG} \\ 
\hline
{Next Release Problem (NRP)} & \cellcolor{BlueCell}0.9244\textsuperscript{+60\%}$^\star$ & \cellcolor{BlueCell}0.8693\textsuperscript{+37\%}$^\dagger$ & \cellcolor{BlueCell}0.9517\textsuperscript{+13\%}$^\dagger$ \\
{Software Configuration Tuning (SCT)} & 0.6932\textsuperscript{+22\%} & \cellcolor{BlueCell}0.7222\textsuperscript{+13\%}$^\star$ & \cellcolor{BlueCell}0.8827\textsuperscript{+5\%}$^\star$ \\
{Software Effort Estimation (SEE)} & \cellcolor{BlueCell}0.7619\textsuperscript{+41\%}$^\star$ & \cellcolor{BlueCell}0.7752\textsuperscript{+23\%}$^\star$ & \cellcolor{BlueCell}0.9056\textsuperscript{+8\%}$^\dagger$ \\
{Software Defect Prediction (SDP)} & \cellcolor{BlueCell}0.8107\textsuperscript{+40\%}$^\dagger$ & \cellcolor{BlueCell}0.8606\textsuperscript{+33\%}$^\dagger$ & \cellcolor{BlueCell}0.9349\textsuperscript{+10\%}$^\dagger$ \\
{Software Product Lines Testing (SPLT)} & \cellcolor{BlueCell}0.7497\textsuperscript{+33\%}$^\star$ & \cellcolor{BlueCell}0.7690\textsuperscript{+20\%}$^\dagger$ & \cellcolor{BlueCell}0.9040\textsuperscript{+7\%}$^\dagger$ \\
{Software Project Scheduling (SPS)} & \cellcolor{BlueCell}0.7690\textsuperscript{+32\%}$^\star$ & \cellcolor{BlueCell}0.8786\textsuperscript{+37\%}$^\dagger$ & \cellcolor{BlueCell}0.9304\textsuperscript{+10\%}$^\dagger$ \\
{Third Party Library Migration (TPLM)} & \cellcolor{BlueCell}0.6988\textsuperscript{+30\%}$^\dagger$ & \cellcolor{BlueCell}0.7864\textsuperscript{+27\%}$^\dagger$ & \cellcolor{BlueCell}0.9002\textsuperscript{+8\%}$^\dagger$ \\
{Workflow Scheduling (WS)} & \cellcolor{BlueCell}0.7429\textsuperscript{+36\%}$^\star$ & \cellcolor{BlueCell}0.7241\textsuperscript{+16\%}$^\star$ & \cellcolor{BlueCell}0.8885\textsuperscript{+6\%}$^\star$ \\
{Web Service Composition (WSC)} & \cellcolor{BlueCell}0.6929\textsuperscript{+27\%}$^\star$ & \cellcolor{BlueCell}0.7259\textsuperscript{+15\%}$^\star$ & \cellcolor{BlueCell}0.8848\textsuperscript{+5\%}$^\star$ \\
\hline
\mytag{R3C1}\rev{\textbf{Average}} & \cellcolor{BlueCell}\rev{\textbf{0.7564}\textsuperscript{+35\%}$^\dagger$} & \cellcolor{BlueCell}\rev{\textbf{0.7929}\textsuperscript{+25\%}$^\dagger$} & \cellcolor{BlueCell}\rev{\textbf{0.9095}\textsuperscript{+8\%}$^\dagger$} \\
\bottomrule
\end{tabular}
\end{adjustbox}
\end{table}

We first compare \approach~with the random guessing using  Wilcoxon signed-rank test.
\mytag{R3C1}\rev{Table~\ref{tab:folder_ndcg_redbetter} shows that \approach~significantly outperforms random ranking in the vast majority of the nine SE problems, achieving stable performance improvements across different settings, with improvements of up to 60\%.} 
Notably, for NDCG@1, the performance is improved on all problems, with eight of them reaching statistical significance. 
\approach~is statistically better on all problems under NDCG@3. 
\mytag{R3C1}\rev{For the full NDCG, statistically significant improvements are also observed on all problems. Notably, \approach~yields improvements of $35\%$, $25\%$, and $8\%$ for NDCG@1, NDCG@3, and full NDCG, respectively.} 

\definecolor{BlueCell}{RGB}{210,230,255}

\begin{table}[t!]
\centering
\scriptsize
\setlength{\tabcolsep}{2.2pt}
\caption{\mytag{R3C4}\rev{Comparing \approach{} with the best top 2 auxiliary-objective choices for all cases/SE problems from our empirical study. The format follows Table~\ref{tb:rq1_result}.}}
\label{tab:fixed_pattern_ndcg_at1}
\resizebox{\columnwidth}{!}{%
\begin{tabular}{lcccccc}
\toprule
\textbf{Problem} &
\multicolumn{2}{c}{\textbf{L2RAC}} &
\multicolumn{2}{c}{\textbf{Top1}} &
\multicolumn{2}{c}{\textbf{Top2}} \\
\cmidrule(lr){2-3} \cmidrule(lr){4-5} \cmidrule(lr){6-7}
& \textbf{Rank} & \textbf{Mean $\pm$ Std}
& \textbf{Rank} & \textbf{Mean $\pm$ Std}
& \textbf{Rank} & \textbf{Mean $\pm$ Std} \\
\midrule
Next Release Problem (NRP) & \cellcolor{BlueCell}1 & \cellcolor{BlueCell}$0.924 \pm 0.209$ & 2 & $0.765 \pm 0.147$ & 3 & $0.739 \pm 0.101$ \\
Software Configuration Tuning (SCT) & \cellcolor{BlueCell}1 & \cellcolor{BlueCell}$0.693 \pm 0.253$ & 2 & $0.570 \pm 0.255$ & 2 & $0.548 \pm 0.226$ \\
Software Effort Estimation (SEE) & \cellcolor{BlueCell}1 & \cellcolor{BlueCell}$0.762 \pm 0.220$ & \cellcolor{BlueCell}1 & \cellcolor{BlueCell}$0.762 \pm 0.200$ & 2 & $0.590 \pm 0.227$ \\
Software Defect Prediction (SDP) & \cellcolor{BlueCell}1 & \cellcolor{BlueCell}$0.811 \pm 0.294$ & \cellcolor{BlueCell}1 & \cellcolor{BlueCell}$0.811 \pm 0.304$ & 2 & $0.775 \pm 0.172$ \\
Software Product Lines Testing (SPLT) & \cellcolor{BlueCell}1 & \cellcolor{BlueCell}$0.750 \pm 0.216$ & 2 & $0.626 \pm 0.214$ & 3 & $0.456 \pm 0.191$ \\
Software Project Scheduling (SPS) & 2 & $0.769 \pm 0.195$ & \cellcolor{BlueCell}1 & \cellcolor{BlueCell}$0.825 \pm 0.169$ & \cellcolor{BlueCell}1 & \cellcolor{BlueCell}$0.849 \pm 0.155$ \\
Third Party Library Migration (TPLM) & \cellcolor{BlueCell}1 & \cellcolor{BlueCell}$0.699 \pm 0.263$ & \cellcolor{BlueCell}1 & \cellcolor{BlueCell}$0.702 \pm 0.250$ & \cellcolor{BlueCell}1 & \cellcolor{BlueCell}$0.739 \pm 0.248$ \\
Workflow Scheduling (WS) & \cellcolor{BlueCell}1 & \cellcolor{BlueCell}$0.743 \pm 0.274$ & 2 & $0.614 \pm 0.374$ & 2 & $0.543 \pm 0.259$ \\
Web Service Composition (WSC) & \cellcolor{BlueCell}1 & \cellcolor{BlueCell}$0.693 \pm 0.265$ & 3 & $0.464 \pm 0.365$ & 2 & $0.557 \pm 0.264$ \\
\midrule[0.8pt]
\rev{\textbf{Average}} & \cellcolor{BlueCell}\rev{\textbf{1.111}} & \cellcolor{BlueCell}\rev{$\mathbf{0.760 \pm 0.243}$} & \rev{1.667} & \rev{$0.682 \pm 0.253$} & \rev{2.000} & \rev{$0.644 \pm 0.205$} \\
\bottomrule
\end{tabular}%
}
\end{table}

\mytag{R3C4}\rev{We further compare \approach{} with the top 2 generally best auxiliary-objective choices derived from empirical average ranks. We use only NDCG@1 since the top 2 methods only have a single choice, and we use the Scott-Knott ESD test for multiple comparisons. As shown in Table~\ref{tab:fixed_pattern_ndcg_at1}, \approach{} obtains the best overall Scott-Knott rank (the best for 8 out of 9 SE problems), with an average NDCG@1 of $0.760 \pm 0.243$, compared with $0.682 \pm 0.253$ for top 1 (4 best) and $0.644 \pm 0.205$ for top 2 (2 best). Thus, the above confirms that simply deriving the patterns observed from our study cannot fulfill the needs of an individual case of the SE problem---yet another piece of evidence for the ``no free lunch theorem''.}

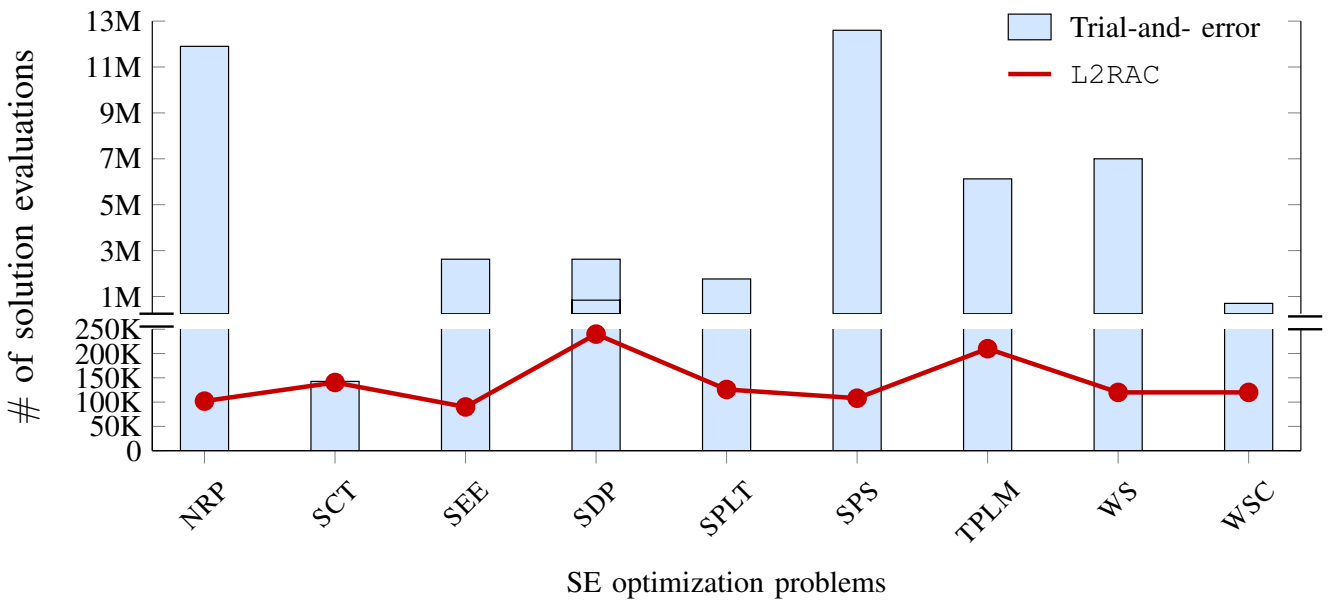
\begin{figure}[t!]
\centering
\begin{tikzpicture}
\begin{axis}[
    name=lower axis,
    height=3cm, width=15cm,
    ymin=0, ymax=250000,
    ytick={0,50000,100000,150000,200000,250000},
    yticklabels={0,50K,100K,150K,200K,250K},
    scaled y ticks=false,
    axis x line*=bottom,
    xtick={1,2,...,9},
    xticklabels={\textsc{NRP},\textsc{SCT},\textsc{SEE},\textsc{SDP},\textsc{SPLT},\textsc{SPS},\textsc{TPLM},\textsc{WS},\textsc{WSC}},
    xticklabel style={rotate=45, font=\small},
    xlabel=SE optimization problems,
    enlarge x limits=0.05,
    ybar=0pt,  
    bar width=16pt,
    grid style=dashed,
    axis on top
]
\addplot[fill=BlueCell, draw=black] coordinates { 
(1,11900000)
(2,142600)
(3,2625000)
(4,840000)
(5,1764000)
(6,12600000)
(7,6125000)
(8,7000000)
(9,700000)
};
\end{axis}

\begin{axis}[
    name=line axis,
    height=3cm, width=15cm,
    ymin=0, ymax=250000,
    axis x line*=none,  
    axis y line*=none,  
    xtick=\empty,
    ytick=\empty,
    xticklabels={},
    yticklabels={},
    enlarge x limits=0.05,
    grid=none  
]
\addplot+[
    mark=*,   
    mark size=2.5pt,  
    color=RedLine,
    line width=1.5pt,
    mark options={fill=RedLine}  
] coordinates { 
(1,102000)
(2,140400)
(3,90000)
(4,240000)
(5,126000)
(6,108000)
(7,210000)
(8,120000)
(9,120000)
};
\end{axis}

\begin{axis}[
    name=upper axis,
    height=5cm, width=15cm,
    at={(0,1.6cm)}, 
    ymin=250000, ymax=13000000,
    ytick={1000000,3000000,5000000,7000000,9000000,11000000,13000000},
    yticklabels={1M,3M,5M,7M,9M,11M,13M},
    scaled y ticks=false,
    axis x line=none,
    xtick={1,2,...,9},
    xticklabels={},
    enlarge x limits=0.05,
    ybar=0pt,
    bar width=16pt,
    grid style=dashed
]
\addplot[fill=BlueCell, draw=black] coordinates { 
(1,11900000)
(4,2625000)
(3,2625000)
(4,840000)
(5,1764000)
(6,12600000)
(7,6125000)
(8,7000000)
(9,700000)
};
\end{axis}

\draw[black, thick] ([shift={(-0.15cm,-0.15cm)}]upper axis.south west) -- ++(0.4cm,0);
\draw[black, thick] ([shift={(-0.15cm,0)}]upper axis.south west) -- ++(0.4cm,0);
\draw[black, thick] ([shift={(-0.15cm,0)}]lower axis.north east) -- ++(0.4cm,0);
\draw[black, thick] ([shift={(-0.15cm,0.15cm)}]lower axis.north east) -- ++(0.4cm,0);

\node[rotate=90] at (-1.5cm, 2.5cm) {\large $\#$ of solution evaluations};

\begin{scope}[xshift=10cm, yshift=4.8cm]
    \draw[fill=BlueCell, draw=black] (0,0) rectangle (0.5,0.3);
    \node[anchor=west] at (0.6,0.15) {Trial-and-
error};
    
    \draw[color=RedLine, line width=1.5pt] (0,-0.4) -- (0.5,-0.4);
    \node[anchor=west] at (0.6,-0.4) {\texttt{L2RAC}};
\end{scope}

\end{tikzpicture}

\caption{The cost in terms of solution evaluations for selecting the right auxiliary objective of MMO between using trial-and-error and predicting via \approach.}


\label{fig:cost}
\end{figure}

\subsection{\mytag{R1C1}\rev{Complexity Analysis}}
\mytag{R1C1}\rev{We now analyze the computational complexity of \approach{} in terms of training and prediction. Let $C$ be the number of training cases; $K$ be the number of candidate auxiliary-objective choices; $n$ be the number of sampled solutions for a case; $d$ be the decision dimension; $m$ be the objective dimension; and $p$ be the number of extracted features. The feature-extraction complexity is dominated by pairwise distance- and dominance-based routines: most distributional and correlation-based features require $\mathcal{O}(nm)$ or $\mathcal{O}(nd)$ time, while distance-based and dominance-based features require $\mathcal{O}(n^2d)$ and $\mathcal{O}(n^2m)$ time in the worst case, respectively. Thus, the feature-preparation complexity is bounded by $\mathcal{O}(CK n^2(d+m))$ for training and $\mathcal{O}(K n^2(d+m))$ for prediction on a new case. Yet, it is worth noting that the training is a one-off process for a particular SE optimization problem. Once the features have been computed, ranking-model prediction over the $K$ candidate auxiliary objectives has an additional complexity of $\mathcal{O}(Kp)$, leading to a total prediction complexity of $\mathcal{O}(Kp)+\mathcal{O}(K n^2(d+m))$.}

\subsection{Cost Saving over Trial-and-error}
\label{sec:cost}

Since \approach~is a pre-trained model, i.e., it is trained once on $n-1$ cases and can be applied to many other cases of a given SE optimization problem, we report on the cost of one training run over $n-1$ cases and its prediction cost for all cases, compared with trial-and-error, as shown in Figure~\ref{fig:cost}. The cost is measured by the number of solution evaluations. The performance labels are reused from our empirical study.
Clearly, for all SE optimization problems, \approach~requires only a small fraction of the evaluation costs needed in trial-and-error: for large-scale problems such as SPS, the relative cost of \approach~is below 1\%; in contrast,
for smaller-scale problems like SCT, the cost savings are less significant, but still lower than the trial-and-error process.


\mytag{R1C1}\rev{Depending on the runtime of solution evaluations in a SE problem, the running-time saving of \approach{} can vary, but it is often significant. For example, for all the cases in WS, trial-and-error, which fully runs MMO over all seven auxiliary-objective choices under the same budget we used in our empirical study, took $\approx58.67$ days. In contrast, \approach~needs $\approx11.53$ days, including one training run and deployment for all cases. If we look at a particular case (since \approach~is pre-trained only once), trial-and-error would still need $269.93$ hours while the deployment in \approach~merely incurs $5.63$ hours---a $97.91\%$ saving on running time.}


\subsection{\rev{Discussion}}

\subsubsection{\mytag{R2C2}\rev{Why \approach{} is Needed}} 

\rev{The post-hoc statistical procedure (i.e., trial-and-error) and \approach{} serve different purposes. The former identifies the best auxiliary objective only after MMO has already been executed under all candidate auxiliary-objective choices. Hence, it is suitable for constructing offline ground truth, but it is not a deployment-time method for choosing the auxiliary objective of a new case. In contrast, \approach{} is used before full-budget MMO runs are conducted. Once trained for a given SE optimization problem, it is capable of automatically ranking the candidate auxiliary-objective choices for a new, unforeseen case using sampled solution data and the extracted features.}

\mytag{R2C2}\rev{This distinction is further supported by the cost discussion in Section~\ref{sec:cost}: once the ranking model has been trained, \approach{} can reuse the learned model for a new case and avoid repeating the full trial-and-error process, while retaining reasonable ranking accuracy according to NDCG.}

\subsubsection{\mytag{R1C2}\rev{On Adaptive Auxiliary Objective}}

\rev{Our goal in this work is to isolate and study the effect of auxiliary-objective choice for MMO before search, which serves as a first step toward this research direction. Indeed, a promising extension is to make MMO adaptive during the search, allowing the auxiliary objective to be dynamically switched or reweighted. Yet, this would also introduce non-trivial challenges, such as:}

\begin{itemize}
    \item \mytag{R1C2}\rev{how to determine when to update the auxiliary objective;}
    \item \mytag{R1C2}\rev{what landscape or population information should be used to justify the update;}
    \item \mytag{R1C2}\rev{and how to avoid unstable search behavior caused by frequent changes.}
\end{itemize}

\mytag{R1C2}\rev{Therefore, we leave this adaptation as future work.}

\section{Threats to Validity}
\label{sec:threats}


\subsection{Construct Validity}


The studied choices of auxiliary objectives are obtained from the literature of general multi-objectivization. To design \approach, we use information theory-based feature selection to define 25 features across three major types, deriving from the theory of MMO, \textbf{Property 3}, and the landscape analysis that represent the characteristics of an optimization problem.

\mytag{R3C2}\rev{For the empirical study, we assess the performance of the optimizers using the value of the target objective, normalized across different cases, which serves as the most common setting in prior work~\cite{harman2001search,harman2012search,DBLP:journals/tse/ChenNKM19,DBLP:journals/tosem/ChenL23}.} As for evaluating \approach, we adopt NDCG, which is a widely used metric for learning to rank cases~\cite{hassan2023uniloc,zhang2018recommending}. We also adopt Scott--Knott ESD~\cite{tantithamthavorn2017mvt} and Wilcoxon signed-rank test for mitigating stochastic bias, both of which are commonly used for SE optimization problems~\cite{DBLP:journals/corr/abs-2507-05995,DBLP:journals/tosem/ChenL23,DBLP:conf/sigsoft/0001L21,DBLP:journals/access/GengYJZLGX18,chen2024adapting}.


However, we agree that the above is not exhaustive and there might be other auxiliary objectives, features, and metrics that could further improve the results. 




\subsection{Internal Validity}

The landscape metrics used for \approach~may need parameterization. For example, when computing local efficient points and Pareto connected components, the number of neighbors $k$ determines the size of the local neighborhood for each solution. \mytag{R2C1}\rev{The hyperparameter settings of MMO, including those of the underlying NSGA-II, follow prior work~\cite{DBLP:journals/tse/XuanJRL12,DBLP:journals/tse/ChenCL24,DBLP:journals/tse/TawosiSPH22,DBLP:journals/jss/NiCWSG19,DBLP:journals/tosem/XiangHLLLY24,DBLP:conf/gecco/ChicanoLNA11,DBLP:conf/evoW/DeshpandeMOS22,DBLP:journals/tpds/ZhuZLL16,DBLP:conf/gecco/0001LY18}. The same principle is applied to the other optimizers and baselines: we use settings that have been reported as generally effective in the corresponding studies~\cite{DBLP:journals/tse/XuanJRL12,DBLP:journals/corr/abs-2507-05995,DBLP:journals/tse/TawosiSPH22,DBLP:journals/apin/MafarjaTATADT23,luo2024lssamplingplus,DBLP:conf/gecco/SabarTS18,DBLP:conf/evoW/DeshpandeMOS22,DBLP:journals/jpdc/AhmadLMAK16,DBLP:journals/cj/JuDH23}.} Indeed, while the above serves as a pragmatic experimental design, we cannot guarantee their optimality.




\subsection{External Validity}

We study nine representative SE optimization problems drawn from diverse software engineering domains. This, to the best of our knowledge, is one of the largest-scale studies of its kind. For each SE optimization problem, we study diverse cases that share different complexity and search spaces, leading to \mytag{R3C1}\rev{212} cases. \mytag{R2C1}\rev{These cover the majority of the cases studied in the literature}, omitting only those that fail to complete in a reasonable time under our available resources. The optimizers chosen in the empirical studies are either the state-of-the-art ones for the SE problems or are closely relevant to the purpose of this work (e.g., MMO\textsubscript{o} and PMO). We have also examined well-known general-purpose optimizers for SBSE, e.g., \texttt{SWAY}. For \approach, we compare it with a random guess and generally top-ranked choices from our empirical results, as we are working on a newly formulated problem. We admit that the above can be further expanded to strengthen the results.

\section{Related Work}
\label{sec:related}

\subsection{Multi-objectivization for SE Optimization Problems}

Multi-objectivization has been widely studied in the optimization community~\cite{DBLP:conf/gecco/BrockhoffFHKNZ07}. Its core idea is to reformulate a single-objective optimization problem into a multi-objective one via auxiliary objectives, altering the structure of the search space. As a naive multi-objectivization, PMO optimize the target objective alongside the auxiliary one as if they were equally important~\cite{lochtefeld2015multi,DBLP:journals/eor/Garza-FabrePR15,DBLP:conf/emo/KnowlesWC01}.  

PMO has been applied to various SE optimization problems\cite{DBLP:conf/ssbse/SoltaniDPDZD18,DBLP:conf/kbse/DerakhshanfarDZ20,DBLP:conf/ssbse/MkaouerKBC14,DBLP:journals/tse/YuanB20}.
For example, in crash reproduction, code coverage and stack-trace similarity have been paired in PMO, aiming to improve the coverage and hence find more complex crashes \cite{DBLP:conf/kbse/DerakhshanfarDZ20}. MMO was originally proposed to cope with the limitation of PMO, such that the auxiliary and target should not be treated as equally important, but rather, the former should only assist the latter. Indeed, MMO has been proven to be effective for software configuration tuning~\cite{DBLP:journals/tse/ChenCL24,DBLP:conf/sigsoft/0001L21}.




However, all the above adopt multi-objectivization for individual SE optimization problems rather than a comprehensive investigation of its suitability to a wide range of SE optimization as a baseline, which is our focus in this work.



\subsection{General-Purpose Optimizers for SE Optimization}

A major research direction in SE domains is the development of \emph{general-purpose optimizers} that can be applied across diverse SE optimization problems~\cite{DBLP:journals/tse/ChenNKM19, DBLP:journals/corr/abs-2506-04509}. The primary goal of these methods is to provide a transferable baseline for cross-domain comparisons, which shares a similar purpose to this work, e.g., \texttt{SWAY}~\cite{DBLP:journals/tse/ChenNKM19}, \texttt{LINE}~\cite{DBLP:journals/corr/abs-2506-04509}, and \texttt{LITE}~\cite{DBLP:journals/corr/abs-2506-04509}.




Yet, the above differ from MMO in several aspects:

\begin{itemize}
    \item They focused on the algorithm level, while MMO lies in an optimization level, which is agnostic to the multi-objective algorithms in theory.
    \item They are designed as the baselines for comparing certain aspects of the optimization in SE problems, e.g., \texttt{SWAY} is for efficiency. None aims to cover the ability to deal with local optima as MMO.
    \item We have demonstrated that, given a suitable auxiliary objective, MMO considerably outperforms all of the above across the SE optimization problems.
\end{itemize}

Further, we propose \approach~as an assist tool to specialize MMO as a strong, yet simplistic baseline for SE optimization. 


\subsection{Empirical Studies on SE Optimization}


There exists a variety of empirical studies for SE optimization under the paradigm of SBSE. For example, in requirements engineering, empirical studies often compare the performance of various optimizers, hoping to recommend the best performing one \cite{del2015multi}. On software configuration tuning, it has been empirically studied how given performance requirements can influence the optimization~\cite{DBLP:journals/tosem/ChenL23a}. General studies for diverse SE optimization problems also exist, e.g., Chen and Li~\cite{DBLP:journals/tosem/ChenL23} reveal that the weighted aggregation can be harmful compared with Pareto optimization, even if clear weights between the objectives are available for three SE optimization problems. Another study~\cite{DBLP:journals/corr/abs-2506-04509} reports the presence of the BINGO effect---only a small fraction of the multi-parameter combinations are effective on five SE optimization problems.

Unlike the above, this work focuses on confirming the effectiveness of MMO on diverse SE problems, paired with theoretical analysis and a predictive tool, which strengthens its suitability as a general baseline for SE optimization.


\section{Conclusion}
\label{sec:conclusion}

This paper studies the possibility of specializing MMO as a general baseline of overcoming local optima for SE optimization problems, particularly relevant to the choice of the auxiliary objectives. Building on the \textbf{Property 3} of MMO, we offer a theoretical analysis to show why and how the auxiliary objective can influence the results. We conduct an empirical study of \mytag{R3C1}\rev{nine SE problems and 212 cases}, showing that if the right auxiliary objective can be used, then MMO:



\begin{itemize}
    \item achieves better results on \mytag{R3C1}\rev{198} cases with considerable improvements than the original MMO and PMO;
    \item generally outperforms the state-of-the-art optimizers for individual SE optimization problem, i.e., with \mytag{R3C1}\rev{167} cases;
    \item remarkably improve the other general-purpose optimizers proposed for the SE domain in \mytag{R3C1}\rev{199} cases.
\end{itemize}

Those theoretical analyses and empirical findings have enabled us to propose \approach, a predictive tool to recommend the best auxiliary objective choice for a given SE optimization problem/case. Experimental results indicate that \approach~consistently outperforms other guessing and such that:

\begin{itemize}
    \item \approach~improves NDCG@1 by \mytag{R3C1}\rev{35\%}, NDCG@3 by \mytag{R3C1}\rev{25\%}, and overall NDCG by \mytag{R3C1}\rev{8\%} across all \mytag{R3C1}\rev{nine} SE problems. All improvements come with statistical significance.

    \item At the case level, overall NDCG is improved in a majority of the cases, with a substantial proportion achieving statistically significant gains; consistent trends are also observed for NDCG@1 and NDCG@3.
\end{itemize}

All of the above contribute to a practical resolution for specializing MMO, given the characteristics of an SE optimization problem/case. In future work, we aim not only to promote MMO toward a ``wider'' scope across more general SE problems, but also make it ``deeper'', addressing inherently multi-objective SE tasks with interactivity~\cite{wang2026conjecture,long2026human}.



\appendix
\subsection{Details of Auxiliary Objective Choices}
\label{app:aux_detail}

For completeness, here we provide the definitions, formulas, and implementation details for each choice of the auxiliary objectives.
\subsubsection{Reciprocal Auxiliary Objectives}

For target objectives with no zero value, the reciprocal of the normalized target objective (within the explored set) can be used as an auxiliary objective~\cite{DBLP:conf/cec/WatanabeS05}. Specifically, the auxiliary objective is defined as:
\begin{equation}
f_a(x_i) = \frac{1}{\tilde{f}_t(x_i)}
\label{eq:reciprocal_obj}
\end{equation}
where \(\tilde{f}_t(x_i)\) denotes the normalized target objective. Note that although this introduces a nonlinear and monotonic correlation between the target and auxiliary objectives, the dynamic changes presented in the explored set, which influences the normalization, can still create uncertainty.

\subsubsection{Gaussian Noise Auxiliary Objectives}

This builds the auxiliary objective by injecting random noise into the objective function, which differs from each evaluation~\cite{DBLP:conf/cec/WatanabeS05}. Specifically, while optimizing the target objective, an auxiliary objective is formed by adding \emph{independently sampled Gaussian noise to each candidate solution} under every evaluation. To ensure comparability between the target and auxiliary objectives, the target objective is normalized (within the an explored set of solutions)~\cite{DBLP:journals/tse/ChenCL24,DBLP:conf/sigsoft/0001L21}.

Formally, this auxiliary noise objective is defined as:
\begin{equation}
f_a(x_i) = \tilde{f}_t(x_i) + \epsilon, \quad
\epsilon \sim \mathcal{N}(0,1)
\label{eq:aux_obj}
\end{equation}
where $\tilde{f}_t(x_i)$ denotes the normalized target objective, $f_a(x_i)$ is the auxiliary objective, and $\epsilon$ is a random variable drawn independently from a standard normal distribution for each solution evaluation.
The auxiliary objective is optimized in the same direction as the target objective. 

\subsubsection{Age Auxiliary Objectives}

This considers the age of each solution explored during optimization as an auxiliary objective~\cite{DBLP:conf/emo/AbbassD03}. Specifically, solutions in the initial set are assigned ages sequentially from one to the solution set size. In subsequent optimization cycles, the age of a solution is defined as the sum of the solution set size and the current iteration count. Thus, the ages of solutions gradually increase as optimization proceeds. Formally, the auxiliary objective is defined as:
\begin{equation}
f_a(x_i) = \text{age}(x_i)
\label{eq:age_obj}
\end{equation}
Naturally, the age auxiliary objective is intended to be maximized to favor older solutions.

\subsubsection{Original Auxiliary Objectives}


This is what has been \rev{designed} with the original MMO~\cite{chen2024adapting,DBLP:journals/tse/ChenCL24,DBLP:conf/sigsoft/0001L21}, i.e., using an extra property of the software artifact as the auxiliary objective (which we do not care about its value), e.g., CPU load of a system when optimizing runtime~\cite{chen2024adapting}; the cost in next release problem when optimizing profit~\cite{DBLP:journals/tse/XuanJRL12,DBLP:journals/ese/DurilloZAHN11}.



\subsubsection{Penalty Auxiliary Objectives}

Penalty mechanisms can improve global search and solution diversity~\cite{DBLP:journals/isci/Alsheddy18}. 
For a solution $x_i$, the auxiliary objective is defined as the sum of the normalized target objective 
and weighted penalties induced by relevant landscape features regarding the local optima:
\begin{equation}
f_a(x_i) = \tilde{f}_t(x_i) + \sum_{l \in L_i} \lambda_l p_l(x_i)
\label{eq:penalty_obj_detailed}
\end{equation}
where $\tilde{f}_t(x_i)$ denotes the normalized target objective, 
$L_i$ is the set of features triggering penalties, 
$\lambda_l$ is a user-defined penalty coefficient controlling the relative influence of the penalty term,
and $p_l(x_i)$ represents the penalty associated with feature $l$. 
In our experiments, $\lambda_l$ is set to $0.1$ for all $l$, which provides a reasonable balance between the target objective and penalty terms without additional fine-grained tuning. 
We note that $\lambda_l$ can be adjusted to reflect different preferences regarding the trade-off between solution quality and constraint violations.
The penalty-based auxiliary objective is minimized, as larger penalties indicate less desirable solutions.

\subsubsection{Grid Diversity Auxiliary Objectives}


Here, the auxiliary objective is formed based on previous grid-based diversity measures \cite{DBLP:journals/tcyb/LiYL14}. This metric characterizes sparsity by computing distances between explored solutions and analyzing their distribution in the grid space:
\begin{equation}
f_a(x_i) = D_{\mathrm{grid}}(x_i, P, \theta)
\label{eq:diversity_obj}
\end{equation}
where $P$ is the current solution set and $\theta \in [0,1]$ is a neighborhood adjustment coefficient controlling the scale of local density computation. The grid diversity auxiliary objective is designed to be maximized to favor solutions in less crowded regions. 

\subsubsection{Novelty Auxiliary Objectives}

Novelty has been introduced as an auxiliary objective to enhance exploration capability in multi-objectivization \cite{DBLP:conf/cec/IshizawaST24,mouret2011novelty,lehman2011novelty}. Specifically, the novelty is often defined as the objective as the mean distance between the a solution and its $k$ nearest neighbors in the explored set:
\begin{equation}
f_a(x_i) = \frac{1}{k} \sum_{j=1}^{k} d(x_i, x_j)
\label{eq:novelty_obj_general}
\end{equation}
where $d(\cdot, \cdot)$ denotes a general distance metric, and $x_j$ represents the $k$ nearest neighbors from the set. The novelty auxiliary objective is intended to be maximized.

\subsection{Overview of SE Optimization Problems and Cases}
\label{app:instances}

Here, we provide an overview of the \mytag{R3C1}\rev{9} SE optimization problems and their cases.


\subsubsection{Next Release Problem}

The problem aims to select a subset of customer requirements for the next software release in order to maximize the expected client profit while respecting all associated constraints. All 17 cases defined in the original study~\cite{DBLP:journals/tse/XuanJRL12} are used in our experiments. Each case specifies a set of customer requirements and the constraints among them. 

The target objective is to maximize client profit, while the auxiliary objective---adopted from the original multi-objective setup---is to minimize development cost.


\subsubsection{Software Configuration Tuning}

The problem aims to improve system performance by adjusting configurable parameters of a software system. Each case corresponds to a specific system together with its configuration space. A solution is represented as a configuration vector, where each variable denotes a tunable option.

We obtained 13 software systems from prior studies~\cite{DBLP:conf/sigsoft/0001L21,DBLP:journals/tse/ChenCL24}, where each system is associated with two performance properties. Following the same procedure, we alternatively optimize each property as the target objective while treating the other as the auxiliary objective in the original MMO formulation, resulting in a total of 26 cases. 


\subsubsection{Software Effort Estimation}
This problem aims to predict the development effort required for software projects based on historical project data. A solution corresponds to a specific configuration of an effort estimation model, with variables controlling aspects such as feature selection and model parameters.

The target objective is to minimize the Sum of Absolute Errors (SAE), which measures the total absolute difference between the actual and predicted effort for all projects in a case. The original auxiliary objective is to minimize the uncertainty of the predicted effort distribution, quantified using the Confidence Interval (CI) of the predictions.

We use the five datasets reported in the original work~\cite{DBLP:journals/tse/TawosiSPH22}. To generate a sufficient number of cases for experimentation and following common practice in effort estimation~\cite{DBLP:journals/tse/TawosiSPH22,DBLP:conf/promise/Azzeh11}, we apply three-fold cross-validation to each dataset. This splits each original dataset into three sub-cases, resulting in a total of 15 cases. 

\subsubsection{Software Defect Prediction}
The problem aims to predict defects in software modules or projects. Each case corresponds to a combination of a dataset and a specific defect prediction model, where a solution represents a selected set of features and model parameter settings.

During the optimization process, objective values are evaluated on the training set, whereas the final results are computed on the test set, following the evaluation protocol of the original work~\cite{DBLP:journals/jss/NiCWSG19}. The target objective is to maximize defect prediction accuracy, and the original auxiliary objective is to minimize the number of selected features.

We adopt the same experimental setup as in the original work~\cite{DBLP:journals/jss/NiCWSG19}, using 10 PROMISE datasets and the same four prediction models: Decision Tree, k-Nearest Neighbors, Logistic Regression, and Naive Bayes. By combining each dataset with each model, a total of 40 cases are generated.

\subsubsection{Software Product Lines Testing}

The problem aims to optimize testing for software product lines by selecting a subset of products to test. Each case corresponds to a software product line instance, where a solution represents a chosen set of products. 

The target objective is to maximize product coverage, and the original auxiliary objective is to minimize the number of selected products.

Due to resource constraints, we select a subset of the datasets provided in the original work~\cite{DBLP:journals/tosem/XiangHLLLY24}, specifically those with fewer than 100 candidate products. This selection results in a total of 21 cases.

\subsubsection{Software Project Scheduling}

The problem aims to optimize scheduling in software projects by assigning tasks to employees in order to meet project objectives. Each case corresponds to a project instance, where a solution represents a specific assignment of employees to tasks.

The target objective is to minimize the overall project duration, and the original auxiliary objective is to minimize the total project cost.

We use 18 cases derived from the same source as the original work~\cite{DBLP:conf/gecco/ChicanoLNA11}, with slight differences in instance configuration, specifically in the number of skills assigned to each employee. The number of tasks is set to 10, 20, or 30, the number of employees is set to five, 10, or 15, and each employee is associated with four--five or six--seven skills.

\subsubsection{Third Party Library Migration}

The problem aims to optimize third-party library migration by selecting API replacements that balance similarity with the original APIs and migration effort. Each case corresponds to a library migration instance, where a solution represents a specific set of API migrations. 

The target objective is to maximize API migration similarity, and the original auxiliary objective is to minimize the number of migrations.

Given the relatively small number of datasets, and following the experimental design in that work~\cite{DBLP:conf/evoW/DeshpandeMOS22}, we consider four different evaluation functions for each dataset. These functions include co-occurrence probability, method similarity, documentation similarity, and a weighted aggregation of the three measures.

By applying these four evaluation functions to each dataset, multiple cases are generated, resulting in a total of 35 cases. One case is excluded because the corresponding single-objective feature cannot be computed in the subsequent prediction task.

\subsubsection{Workflow Scheduling}

The problem focuses on workflow scheduling, where the goal is to allocate tasks to resources to optimize execution. Each case corresponds to a workflow instance, and a solution represents a specific assignment of tasks to resources.

The target objective is to minimize the makespan of the workflow execution, while the original auxiliary objective is to minimize the total execution cost.

We use the same 20 cases as in the original work~\cite{DBLP:journals/tpds/ZhuZLL16}, with each dataset corresponding to one case, resulting in a total of 20 cases.

\subsubsection{Web Service Composition}

The problem focuses on Web Service Composition, where the goal is to combine multiple web services to meet user requirements while optimizing quality metrics. Each case corresponds to a set of candidate services, and a solution represents a specific composition of these services. 

The target objective is to maximize throughput, while the original auxiliary objective is to minimize latency.

Following the procedure in previous works~\cite{DBLP:conf/gecco/0001LY18,zheng2012investigating}, we independently generate 10 datasets using the same methodology. To avoid favoring any specific Quality-of-Service (QoS) metric, we alternately treat throughput and latency as the target and auxiliary objectives. This process results in a total of 20 cases.

\subsection{Details of the Features for \approach}
\label{app:feature_definitions}


Here, we elaborate the details of the three categories of features used in \approach.


\subsubsection{Signle objective Landscape Features}
\textbf{Skewness (\texttt{Ske})} \cite{pearson1895contributions} and \textbf{Kurtosis (\texttt{Kur})} \cite{decarlo1997meaning} are statistical measures that describe the shape of a distribution of solutions from different perspectives. Specifically, \texttt{Ske} quantifies the asymmetry of the distribution: a positive skewness indicates a longer right tail, while a negative skewness indicates a longer left tail, providing insights into the imbalance among solutions. \texttt{Kur} measures the peakedness or flatness of the distribution, where a value greater than three indicates a more peaked distribution and a value less than three suggests a flatter distribution, reflecting the concentration and spread of solutions. Formally, they are calculated as:
\begin{equation}
Ske = \frac{\mathbb{E}[(Y - \mu)^3]}{\sigma^3}
\end{equation}
\begin{equation}
Kur = \frac{\mathbb{E}[(Y - \mu)^4]}{\sigma^4}
\end{equation}
where $Y$ denotes the values of solutions, and $\mu$ and $\sigma$ represent the mean and standard deviation, respectively. Larger absolute values of \texttt{Ske} indicate greater asymmetry, while higher \texttt{Kur} values correspond to more concentrated or peaked solution distributions.

\textbf{Maximum Information Entropy (\texttt{MIE})} \cite{6719480} quantifies the information complexity of a distribution across the solution space.
\begin{align}
MIE &= \max_{\epsilon} H(\epsilon) \\
H(\epsilon) &= -\sum_{i,j} p_{ij}(\epsilon) \log_6 p_{ij}(\epsilon)
\end{align}
where $p_{ij}(\epsilon)$ represents the transition probabilities in an $\epsilon$-based discretization of the solution space. \texttt{MIE} reflects the heterogeneity and unpredictability of the landscape: higher \texttt{MIE} values indicate more complex structures and greater variability among solutions, while lower values suggest a more uniform or regular distribution.

\textbf{Nearest Better Clustering (\texttt{NBC})} \cite{10.1145/2739480.2754642} measures the relative clustering of solutions in the solution space.
\begin{equation}
NBC = \frac{1}{|X|} \sum_{x \in X} \frac{d_{nb}(x)}{d_{nn}(x)}
\end{equation}
where $d_{nb}(x)$ is the distance to the nearest better solution according to the chosen metric, and $d_{nn}(x)$ is the distance to the nearest neighbor solution. Higher \texttt{NBC} values indicate that better solutions are more closely clustered, suggesting localized concentration, while lower values imply more dispersed or scattered solution structures.

\subsubsection{Dominance-based Features}

\textbf{Pareto Dominated Solution Ratio (\texttt{PDSR})} \cite{DBLP:conf/evoW/VerelLJD11} quantifies the proportion of solutions that are dominated by at least one other solution:

\begin{equation}
PDSR = \frac{|\{x_i \in X \mid \exists\, x_j \in X, \; x_j \prec x_i\}|}{|X|}
\end{equation}
where $x_j \prec x_i$ denotes that $x_j$ Pareto-dominates $x_i$. The value of \texttt{PDSR} ranges from 0 to 1. Higher values indicate that most solutions are dominated, implying a steep dominance hierarchy, while lower values suggest a more balanced distribution in which only a few solutions dominate others.

\textbf{Level Count (\texttt{LC})} \cite{DBLP:conf/ppsn/LiangCL24} quantifies the number of non-dominated fronts obtained by stratifying the solution set $X$ using non-dominated sorting. In this procedure, Level one contains globally non-dominated (Pareto-optimal) solutions, Level two consists of non-dominated solutions after removing Level one, and the process continues iteratively until all solutions are assigned to a dominance level:
\begin{equation}
LC = \text{Number of non-dominated fronts in } X
\end{equation}
where higher \texttt{LC} values correspond to deeper dominance hierarchies in the Pareto landscape, indicating a more complex dominance structure with multiple intermediate layers between inferior and superior solutions.

\textbf{Local Efficient Points Count (\texttt{LEPC})} \cite{DBLP:conf/ppsn/LiangCL24,DBLP:conf/ppsn/KerschkeWPGDTE16} quantifies the number of locally efficient solutions within the solution set $X$ of multi-objective optimization problems. A solution $x_i \in X$ is locally efficient if no neighbor $x_j \in \mathcal{N}(x_i)$ Pareto-dominates it. Formally:
\begin{equation}
LEPC = |\{x_i \in X \mid \nexists\, x_j \in \mathcal{N}(x_i), \; x_j \prec x_i\}|
\end{equation}
where higher \texttt{LEPC} values imply the existence of multiple locally optimal regions, indicating stronger multimodality and smoother local dominance structures.

\textbf{Local-to-Global Ratio (\texttt{LGR})} quantifies the relationship between Pareto local optimal and global Pareto-optimal solutions. While existing studies often analyze local and global optima separately \cite{liefooghe2019landscape,verel2013structure}, \texttt{LGR} provides a unified measure to quantify their alignment, offering crucial insights into landscape characteristics. It is computed as the proportion of Pareto local optima that are also globally Pareto-optimal:
\begin{equation}
\texttt{LGR} = \frac{|X_{\text{local}} \cap X_{\text{global}}|}{|X_{\text{local}}|}
\end{equation}
where $X_{\text{local}}$ denotes the set of Pareto local optimal solutions (non-dominated within their neighborhoods), and $X_{\text{global}}$ represents the set of globally Pareto-optimal solutions. Higher \texttt{LGR} values indicate smoother landscapes where local optima coincide with global optima, while lower values reflect rugged search spaces with deceptive local structures that may mislead local search procedures.

\textbf{Connected Components (\texttt{CC})} \cite{DBLP:conf/ppsn/KerschkeWPGDTE16} measures the connectivity among locally efficient solutions in the solution space. This feature quantifies the number of disjoint, maximally connected subsets formed by these solutions. Formally, \texttt{CC} is computed from a graph $G_L$, where each vertex represents a locally efficient solution, and edges connect two vertices if their corresponding solutions are spatially adjacent:
\begin{equation}
CC = \text{Number of connected components in } G_L
\end{equation}
Higher \texttt{CC} values indicate that locally efficient regions are more scattered and less connected, reflecting fragmented dominance structures.

\textbf{Connected Components (Pareto) (\texttt{CC-P})} \cite{DBLP:conf/ae/LiefoogheVAT13} focuses on the connectivity among globally Pareto-optimal solutions. It is computed from a Pareto graph $G_P = (V, E)$, where vertices correspond to Pareto-optimal solutions, and edges are defined based on a spatial neighborhood relation:
\begin{equation}
CC\text{-}P = \text{Number of connected components in } G_P
\end{equation}
Larger \texttt{CC-P} values indicate that Pareto-optimal solutions are divided into multiple disconnected clusters, suggesting a fragmented Pareto structure.

\textbf{Maximum Dominance Count (\texttt{MDC})}, \textbf{Median Dominance Count (\texttt{MedDC})}, \textbf{Dominance Change Frequency (\texttt{DCF})}, and \textbf{Average Dominance Pressure (\texttt{ADP})} are dominance-based features that characterize the dominance structure in multi-objective optimization from different perspectives. The Pareto dominance relation serves as the fundamental principle for comparing solutions~\cite{liefooghe2019landscape,DBLP:conf/gecco/LiefoogheVLZM21}. Formally, they are defined as:
\begin{equation}
\texttt{MDC} = \max_{i \in [1,n]} \left|\{j \in [1,n] \mid \mathbf{x}_i \prec \mathbf{x}_j\}\right|
\end{equation}
\begin{equation}
\texttt{MedDC} = \text{median}_{i \in [1,n]} \left|\{j \in [1,n] \mid \mathbf{x}_i \prec \mathbf{x}_j\}\right|
\end{equation}
\begin{equation}
\texttt{DCF} = \frac{1}{n} \sum_{i=1}^{n} \frac{\left|\{j \in \mathcal{N}(i) \mid \text{dom}(\mathbf{x}_i) \neq \text{dom}(\mathbf{x}_j)\}\right|}{|\mathcal{N}(i)|}
\end{equation}
\begin{equation}
\texttt{ADP} = \frac{1}{n} \sum_{i=1}^{n} \frac{\left|\{j \in \mathcal{N}(i) \mid \mathbf{x}_j \prec \mathbf{x}_i\}\right|}{|\mathcal{N}(i)|}
\end{equation}
Larger \texttt{MDC} values indicate highly dominant solutions and steep dominance gradients; higher \texttt{MedDC} represents stronger overall dominance; larger \texttt{DCF} suggests irregular and rugged dominance transitions; and higher \texttt{ADP} reflects stronger local dominance pressure among solutions.

\textbf{Hypervolume (\texttt{HV})}~\cite{DBLP:journals/tec/ZitzlerTLFF03} quantifies the portion of the objective space weakly dominated by an approximation set. For minimization problems, given a reference point $\mathbf{r} = (r_1, \dots, r_m)$ bounding the objective space:
\begin{equation}
HV(A, \mathbf{r}) = \text{vol}\left( \bigcup_{\mathbf{z} \in A} [z_1, r_1] \times \cdots \times [z_m, r_m] \right)
\end{equation}
A larger \texttt{HV} indicates better convergence, diversity, and strict Pareto compliance.

\textbf{Pareto Grid Ratio (\texttt{PGR})} measures the spatial extent and dispersion of Pareto-optimal solutions in the objective space. The objective space is discretized into uniform grid cells, where $N_{\text{occupied}}$ is the number of cells containing at least one Pareto-optimal solution, and $N_{\text{total}}$ is the total number of grid cells:
\begin{equation}
PGR = \frac{N_{\text{occupied}}}{N_{\text{total}}}
\end{equation}
Higher \texttt{PGR} values indicate that Pareto-optimal solutions are more widely distributed, reflecting a broader and more diverse Pareto front, while lower values suggest concentration within limited objective regions. This feature is part of a broader family of grid-based descriptors, including \texttt{GDV} and \texttt{AGDM} introduced in subsequent features.

\subsubsection{Objective-space Distribution Features}

\textbf{Spearman Correlation Coefficient (\texttt{SCC})}~\cite{DBLP:conf/gecco/LiefoogheL23} measures the monotonic correlation between pairs of objectives. Formally:
\begin{equation}
\texttt{SCC} = 1 - \frac{6 \sum_{i=1}^{n} d_i^2}{n(n^2 - 1)}
\end{equation}
where $d_i$ denotes the difference between the ranks of the $i$-th solution under two objectives, and $n$ is the total number of solutions. Higher positive values indicate strong positive correlation; lower negative values indicate strong negative correlation; values near zero suggest weak or no correlation.

\textbf{Pearson Correlation Coefficient (\texttt{PCC})} evaluates the linear relationship between pairs of objectives:
\begin{equation}
PCC = \frac{\sum_i (f_1(x_i) - \bar{f}_1)(f_2(x_i) - \bar{f}_2)}{\sqrt{\sum_i (f_1(x_i) - \bar{f}_1)^2 \sum_i (f_2(x_i) - \bar{f}_2)^2}}
\end{equation}
High positive or negative \texttt{PCC} indicates that objectives are closely aligned or strongly conflicting, respectively.

\textbf{Kendall's Tau Correlation Coefficient (\texttt{KTC})} evaluates the ordinal association between objectives using concordant and discordant pairs:
\begin{equation}
KTC = \frac{C - D}{\frac{1}{2} n(n-1)}
\end{equation}
where $C$ and $D$ denote the number of concordant and discordant pairs, respectively. \texttt{KTC} is robust to outliers and effectively captures rank consistency between objectives in complex landscapes.

\textbf{Slope of Linear Regression Fit (\texttt{SLR})} quantifies the regression trend between two objectives by fitting a linear model:
\begin{equation}
f_2(x_i) = \alpha + \beta f_1(x_i) + \epsilon_i
\end{equation}
where $\beta$ denotes the estimated slope coefficient. The magnitude of \texttt{SLR} indicates the sensitivity of one objective to changes in another, providing insights into trade-off intensity and search directionality.

\textbf{Objective Function Value Variance (\texttt{OFV})}~\cite{DBLP:journals/tec/MunozKH15} quantifies the dispersion of objective values across the solution set. Originally proposed for single-objective optimization, it is extended here to multi-objective problems by computing the variance for each objective separately and analyzing their combined behavior:

\begin{equation}
OFV = \frac{1}{n} \sum_{i=1}^{n} (f(x_i) - \bar{f})^2
\end{equation}

Higher \texttt{OFV} values indicate greater variability among objective values, reflecting a more heterogeneous or rugged objective-space landscape.

\textbf{Objective Function Value Entropy (\texttt{OFE})}~\cite{liefooghe2019landscape} quantifies the distributional complexity of objective values using information entropy:
\begin{equation}
OFE = - \sum_{k=1}^{K} p_k \log p_k
\end{equation}
where $p_k$ denotes the normalized frequency of objective values in bin $k$. Entropy is computed for each objective and averaged across all objectives. Higher \texttt{OFE} values imply a more uniform and thus more complex or rugged distribution of objective values.

\textbf{Average Distance (\texttt{AD})}~\cite{DBLP:conf/ae/LiefoogheVAT13} quantifies the mean pairwise dissimilarity among non-dominated solutions in the Pareto set, serving as a core indicator of spatial diversity:
\begin{equation}
AD = \frac{2}{n(n-1)} \sum_{i < j} d(x_i, x_j)
\end{equation}
where $n$ denotes the number of non-dominated solutions in the Pareto set, and $d(\cdot,\cdot)$ represents the distance measure in the solution space. Higher \texttt{AD} values indicate that Pareto-optimal solutions are more widely dispersed, reflecting stronger diversity and potentially higher exploration costs for neighborhood-based search algorithms.

\textbf{Grid Density Variance (\texttt{GDV})} evaluates the non-uniformity of Pareto-optimal solution distribution across the discretized objective space. The objective space is uniformly divided into $G$ grid cells, and $n_g$ denotes the number of solutions within each grid cell:
\begin{equation}
GDV = \frac{1}{G} \sum_{g=1}^{G} (n_g - \bar{n})^2
\end{equation}
where $\bar{n}$ is the average number of solutions per grid cell. Higher \texttt{GDV} values indicate uneven occupancy across the objective space, suggesting irregular convergence patterns or localized clustering of Pareto-optimal solutions. Together with \texttt{PGR} and \texttt{AGDM}, \texttt{GDV} contributes to a unified grid-based characterization of objective-space structure and diversity.

\textbf{Adjacent Grid Distance Mean (\texttt{AGDM})} characterizes the spatial smoothness between neighboring regions in the discretized objective space. Let $\mathcal{E}$ denote the set of edges connecting adjacent grid cells, and $d(\cdot,\cdot)$ represent the distance measure in the objective space. The feature is defined as:
\begin{equation}
AGDM = \frac{1}{E} \sum_{(g_i, g_j) \in \mathcal{E}} d(g_i, g_j)
\end{equation}
where $E = |\mathcal{E}|$ is the number of adjacent grid pairs. Larger \texttt{AGDM} values indicate smoother transitions between neighboring grid regions, reflecting gradual variation in objective values; smaller values suggest abrupt structural changes or fragmented distributions in the Pareto front. \texttt{AGDM} complements \texttt{PGR} and \texttt{GDV} by emphasizing local adjacency smoothness within the same grid-based framework.



\bibliographystyle{IEEEtran}
\bibliography{IEEEabrv,reference}

\begingroup
\makeatletter
\def\@IEEEtranneedspace#1#2{}
\expandafter\patchcmd\csname\string\IEEEbiography\endcsname
  {\vskip \@IEEEBIOskipN plus 1fil minus 0\baselineskip}
  {\vskip 1.8\baselineskip}{}{}
\patchcmd{\IEEEbiographynophoto}
  {\vskip 4\baselineskip plus 1fil minus 0\baselineskip}
  {\vskip 1.8\baselineskip}{}{}
\makeatother

\begin{IEEEbiography}[{\includegraphics[width=1in,height=1.25in,clip,keepaspectratio]{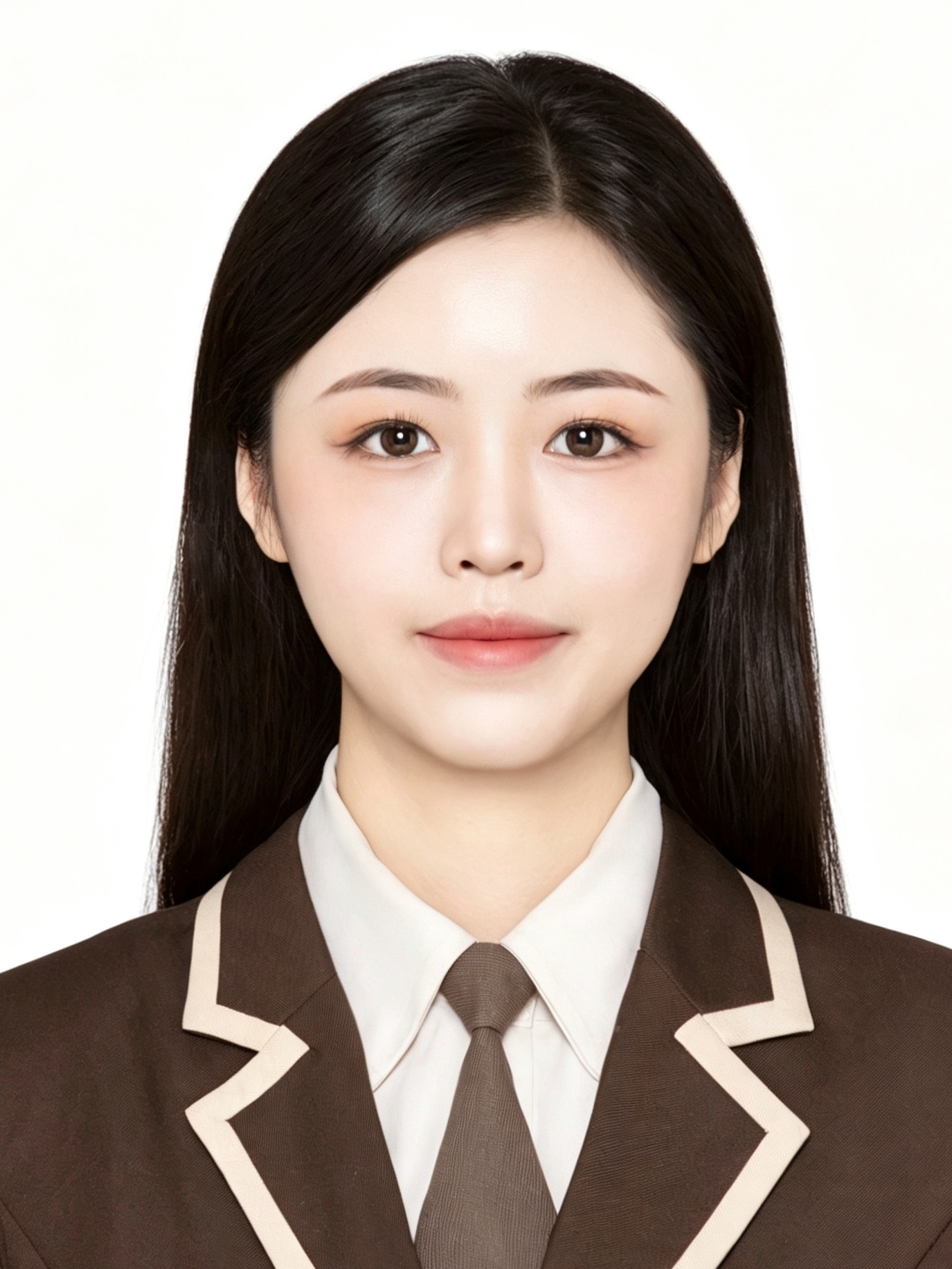}}]{Chengjie Chuan}
is currently a Graduate Student with the School of Computer Science and Engineering, University of Electronic Science and Technology of China. Her research interests include search-based software engineering, multi-objective optimization, and large language models.
\end{IEEEbiography}

\begin{IEEEbiography}
[{\includegraphics[width=1in,height=1.25in,clip,keepaspectratio]{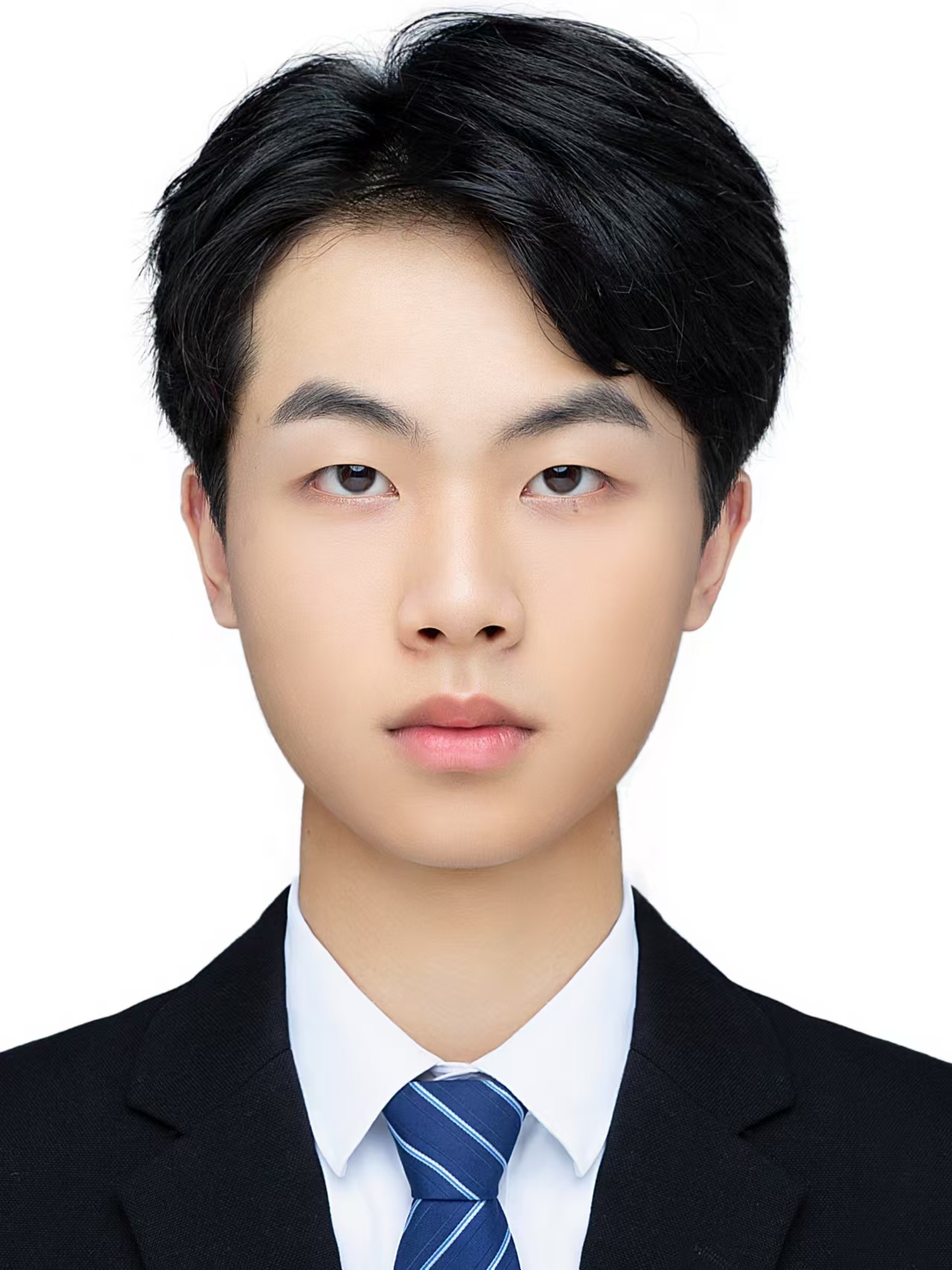}}]{Pengzhou Chen} received the B.Eng. degree in Computer Science from the University of Electronic Science and Technology of China. As a member of IDEAS Lab, he is currently working towards the M.Sc. degree, with the School of Computer Science and Engineering, University of Electronic Science and Technology of China. His research interests include search-based software engineering, performance modeling, and machine learning.
\end{IEEEbiography}

\begin{IEEEbiography}[{\includegraphics[width=1in,height=1.25in,clip,keepaspectratio]{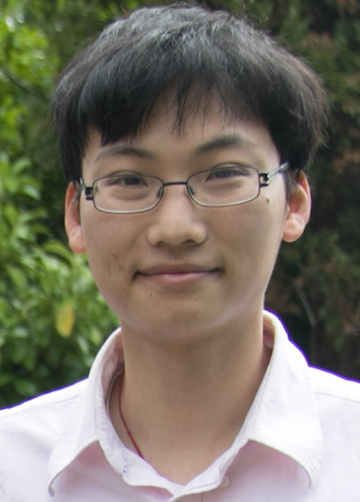}}]{Tao Chen} directs the Intelligent Dependability Engineering for Adaptive Software Laboratory (IDEAS Lab), conducting cutting-edge research on the intersection between AI and Software Engineering. Currently, his research interests include performance engineering, self-adaptive systems, search-based software engineering, data-driven software engineering, and their interplay with machine learning and computational intelligence. His work has been regularly published in all major Software Engineering and AI conferences/journals (e.g., ICSE, FSE, ASE, IJCAI, ACL, CVPR, TOSEM, and TSE), won ACM Distinguished Paper Awards, and has been supported by projects worth over $\pounds$1 million from external funding bodies. He currently serves as an Associate Editor for ACM Transactions on Autonomous and Adaptive Systems and a Program Committee member for many prestigious conferences.
\end{IEEEbiography}

\endgroup

\end{document}